\documentclass[a4paper,10pt]{article}
\usepackage[dvips]{graphicx}
\usepackage{latexsym}
\usepackage{amsmath,amssymb,epsfig,float}
\newcommand{\nn}{\nonumber}
\newcommand{\be}{\begin{equation}}
\newcommand{\ee}{\end{equation}}
\newcommand{\ba}{\begin{array}}
\newcommand{\bqa}{\begin{eqnarray}}
\newcommand{\eqa}{\end{eqnarray}}
\newcommand{\cO}{{\cal O}}

\begin{document}
\title{ \bf \boldmath Meson-baryon reactions with strangeness --1  within a chiral framework }
  \author{Zhi-Hui~Guo$^{a,b}$\thanks{guo@um.es}, J.~A.~Oller$^{b}$\thanks{oller@um.es}  
  \vspace{0.3cm} 
\\ {\footnotesize $^a$ Department of Physics, Hebei Normal University, 050024 Shijiazhuang, P.~R.~China. }
  \\ {\footnotesize $^b$ Departamento de F\'isica, Universidad de Murcia,  E-30071 Murcia, Spain. }
 }

\date{\today}
\maketitle

\begin{abstract}
We study  meson-baryon scattering with strangeness --1 in  unitary chiral perturbation theory. 
Ten coupled channels are considered in our work, namely $\pi^0 \Lambda$, $\pi^0 \Sigma^0$, $\pi^- \Sigma^+$
, $\pi^+ \Sigma^-$, $K^- p$, $\bar{K}^0 n$, $\eta \Lambda$, $\eta \Sigma^0$, $K^0 \Xi^0$ and $K^+ \Xi^-$. 
A large amount of experimental data are analyzed, including the recent precise measurement by the SIDDHARTA Collaboration 
of the energy shift and width of the $1s$ state of kaonic hydrogen. This leads to a strong constraint on the free parameters   
in our theory and of the resulting meson-baryon scattering amplitudes. We also analyze the uncertainty that stems by using several different strategies to perform the fits to data. It is found that large uncertainties 
in the subthreshold extrapolation of the $K^-p$ scattering amplitude 
arise by employing either only one common weak pseudoscalar decay constant or distinguishing between 
$f_\pi$, $f_K$ and $f_\eta$. However, in both cases a good reproduction of experimental data is obtained. 
 We  also discuss the pole content of the resulting $S$-wave amplitudes, particularly in connection with the two-pole structure of the $\Lambda(1405)$ resonance.
\end{abstract}

\\
\noindent{ {\bf Keywords:}  Meson-baryon scattering with strangeness. Chiral perturbation theory. Resonances

\newpage

\section{Introduction}
\label{intro}

The antikaon-nucleon reaction is quite an interesting subject in hadron physics. At low energies Chiral Perturbation Theory (ChPT) \cite{weinberg79,gasser84,gasser85} 
constrains strongly the possible interactions but at the same time strong nonperturbative effects take place. In this respect we 
have the presence of the $\Lambda(1405)$ resonance 
between the $\pi\Sigma$ and $\bar{K}N$ thresholds, which is also a manifestation of strong coupled-channel dynamics. There are also big 
$SU(3)$ symmetry breaking effects, giving rise to a large split in thresholds because of the different masses of the coupled channels composed by the lightest octets of 
pseudoscalars and baryons. Nevertheless, in the chiral limit all these channels are degenerate which implies that despite the potential
presence of large differences in the thresholds all of them are relevant in a wide range of energies. By the same reason a realistic study of the resonances in meson-baryon scattering with strangeness $-1$  in the energy region between 1 and 2~GeV requires to take all of them into account. 
Within $SU(3)$ baryon ChPT no explicit (baryonic or mesonic) resonances are introduced in the Lagrangians. 
In particular, we do not include the $\frac{3}{2}^+$ decuplet of baryon resonances. Since in our present study we are concerned with $S$-wave scattering near 
thresholds, the effects of these resonances should be properly encoded in the low energy counterterms of the ChPT Lagrangians. 
However, for $P$-waves the $\Sigma(1385)$ is below the $\bar{K}N$ threshold and it should be explicitly included, or generated by a proper  multichannel unitarization approach, like already done for the scalar and vector resonances in meson-meson scattering \cite{iam}.

 In recent years, great progress by exploiting  chiral effective field theory 
has been made on this research field
~\cite{kaiser:1995,Oset:1997it,Oller:2000fj,Oller:2005ig,Oller:2006jw,GarciaRecio:2002td,Jido:2003cb,Borasoy:2004kk,Borasoy:2005ie,Magas:2005vu,Hyodo:2002pk,Ikeda:2011pi,Ikeda:2012au,Cieply:2011nq,Mai:2012dt}. These studies corroborate the nature of the $ \Lambda(1405)$ \cite{dalitz} as a dynamically generated resonance from $\bar{K}N$ and $\pi\Sigma$ interactions. However, in Ref.~\cite{Oller:2000fj} it was already noticed the presence of indeed two nearby poles in the $\Lambda(1405)$ 
region. This fact concerning the two-pole structure of the $\Lambda(1405)$  has been studied in subsequent papers ~\cite{Oller:2005ig,Oller:2006jw,GarciaRecio:2002td,Jido:2003cb,Borasoy:2005ie,Magas:2005vu,Ikeda:2011pi,Ikeda:2012au,Cieply:2011nq,Shevchenko:2011ce}. The two poles are commonly characterized as following: the first pole in the complex energy plane is located quite close to 
the $\bar{K}N$ threshold with a small imaginary part, around 10-30~MeV, and a strong coupling to $\bar{K}N$. In turn, 
  the second one is wider,  with a relatively large imaginary part around 50-200 MeV, coupling more strongly to the $\pi\Sigma$ channel and its pole position shows more dependence on the  specific theoretical model \cite{Jido:2003cb,Borasoy:2005ie}.

On the other hand, the $K^-p$ scattering length can be determined by the measurement of the energy shift and width of the $1s$ state in kaonic hydrogen. 
Due to the controversial results of the DEAR Collaboration \cite{dear} and the low precision of the KEK measurement \cite{kek} 
there was an uncertainty of around a factor 2 in this scattering length \cite{Oller:2005ig,Oller:2006jw,Borasoy:2004kk}. 
In this respect the recent precise measurement by the SIDDHARTA Collaboration \cite{Bazzi:2011zj} may finally fix the $K^-p$ scattering length, 
up to a precision of around 20\%. 
Of course, an important point now is to reproduce this measurement simultaneously with  all the other scattering data. 
Studies in this direction are already accomplished in Refs.~\cite{Ikeda:2011pi,Ikeda:2012au,Cieply:2011nq,Mai:2012dt}. 
In the present work we give a step forward and take into account all the  data 
included in Refs.~\cite{Ikeda:2011pi,Ikeda:2012au,Cieply:2011nq,Mai:2012dt} and also consider additional data in our fits, 
such as the $\pi^-\Sigma^+$ event distribution \cite{hemmingway85npb}, 
the cross section of $K^- p \to \eta \Lambda$ \cite{starostin01prc}, 
measurements from the reaction of $K^- p \to \pi^0\pi^0\Sigma^0$ \cite{prakhov04prc} and 
the $\pi \Lambda$ phase shift at the $\Xi^-$ mass \cite{huang,e756}, $\delta_{\pi\Lambda}$.  
Moreover, to further constrain the fits we include  the pion-nucleon isospin even $S$-wave scattering length $a^+_{0+}$, 
the nucleon $\sigma$ term $\sigma_{\pi N}$ and the masses of the $N$, $\Lambda$, $\Sigma$ and $\Xi$. 
We do not aim to give a precise description for $a^+_{0+}$, $\sigma_{\pi N}$ and the masses, 
since we only calculate them up to $\cO(p^2)$ at tree level in $SU(3)$ baryon ChPT.  
Potentially large corrections from loops and higher order low energy constants could exist. 
 So in the fit we allow relatively large errors for those quantities in order to account for 
our theoretical uncertainties. Even so, we still find that they can lead to important constraints on the free parameters. In addition, 
we discuss here two sources of ambiguity overlooked in previous studies. One point concerns with using either just one common weak decay constant 
for all the channels or distinguishing between $f_\pi$, $f_K$ and $f_\eta$. 
The other controversial issue that also requires a closer look
 is the definition employed for the $\chi^2$, as it has become customary in many cases to weight differently 
the data fitted corresponding to different  observables. The first point is of relevance because 
it shows that the uncertainty affecting the subthreshold extrapolation of the $K^-p$ elastic $S$-wave amplitude is much larger than estimated in the recent studies \cite{Ikeda:2011pi,Ikeda:2012au,Cieply:2011nq,Mai:2012dt} that also reproduced the SIDDHARTA measurement. Let us recall that this subthreshold extrapolation is of great interest for the active and controversial field on $\bar{K}$ few-nucleon systems and possible existence of bound states of $\bar{K}$ with heavy nuclei \cite{old1,old2,old3,Cieply:2011yz}.

As an output of our study, we can also obtain important information on the rich resonant content or spectroscopy associated to strangeness $-1$ meson-baryon $S$-wave scattering. 
In this respect, for $I=0$ we confirm the two-pole nature for the $\Lambda(1405)$ resonance and find poles associated with the resonance $\Lambda(1670)$.  In $I=1$ we find poles that are more dependent on the details of the fits. Poles in the region of the $\Sigma(1620)$ are found. 
Other poles corresponding to the $\Sigma(1750)$ and to  more controversial $I=1$ resonances around the $\bar{K}N$ threshold 
  are obtained in one of the fits.

After this introduction, we present the basic set up of our approach in Sec.~\ref{fromalism}. Next, we discuss in Sec.~\ref{fitresult} the fits to the data included with the interacting kernel fixed at leading and next-to-leading order in the chiral expansion. We find a good reproduction of scattering data together with the energy shift and width of the $1s$ state of kaonic hydrogen.  Finally, we discuss in Sec.~\ref{poles} the pole content of our solutions. Section~\ref{conclusions} includes a summary of the work and our conclusions. 

\section{Formalism}
\label{fromalism}

Our starting point is the $SU(3)$ meson-baryon chiral Lagrangian at $\cO(p)$ and $\cO(p^2)$ \cite{Oller:2006yh}: 
\begin{eqnarray}
{\cal L}_1&=&\langle i\bar{B}\gamma^\mu [D_\mu,B]\rangle-m_0 
\langle\bar{B}B \rangle  +\frac{D}{2}\langle \bar{B}\gamma^\mu
\gamma_5\{u_\mu,B\}\rangle +\frac{F}{2} \langle \bar{B}\gamma^\mu 
\gamma_5 [u_\mu,B]\rangle \,,  \nn \\
{\cal L}_2&=&b_0\langle \bar{B}B\rangle \langle 
\chi_+\rangle + b_D\langle{\bar{B}\{\chi_+,B\}}\rangle + b_F \langle
\bar{B}[\chi_+,B] + b_1\langle \bar{B}[u_\mu,[u^\mu,B]]\rangle \nn\\
&+&  b_2 \langle
\bar{B}\{u_\mu,\{u^\mu,B\}\} \rangle + b_3\langle \bar{B}\{u_\mu,[u^\mu,B]\}\rangle
+b_4\langle \bar{B} B\rangle \langle u_\mu u^\mu\rangle+\cdots \,,
\label{lagp2}
\end{eqnarray}
where we do not show those terms, represented by the ellipsis, that do not contribute to the $S$-wave meson-baryon scattering at tree level. 
The octet of baryons $N, \Lambda, \Sigma$ and $\Xi$ is collected in the unitary $3 \times 3$ matrix $B$ and the octet of light pseudoscalar mesons $\pi, K$, and $\eta$ 
is introduced through the basic chiral building blocks: the covariant derivative $D_\mu$, $\chi_+$ and $u_\mu$. 
In our notation, $u_\mu = i u^\dagger \partial_\mu U u^\dagger $, $U=u^2=e^{i \sqrt{2}\Phi/f}$ 
with $\Phi$ the unitary $3 \times 3$ matrix for the octet of light pseudoscalars $\pi, K$, and $\eta$. 
In this way the parameter $f$ corresponds to the pseudoscalar weak decay constant $f$ in the $SU(3)$ chiral limit \cite{Pich:1995bw}. 
Furthermore, $D_\mu B=\partial_\mu B + [\Gamma_\mu, B]$ with $\Gamma_\mu= (u^\dagger \partial_\mu u + u\partial_\mu u^\dagger)/2$ and 
$\chi_+= 2 B_0 (u^\dagger M_q  u^\dagger + u M_q  u)$ with $M_q=\{m_u,m_d,m_s\}$ the diagonal mass matrix 
for the three light flavor quarks. $B_0$ is defined in terms of the vacuum quark condensate in the $SU(3)$ chiral limit
 as $\langle 0|\bar{q}^i q^j |0\rangle=-f^2 B_0 \delta^{ij}$ \cite{Pich:1995bw}. 
The  $\cO(p^2)$ low energy constants $b_0, b_D, b_F, b_{i=1,2,3,4}$  will be fitted 
to data. The values for $D$ and $F$ are taken from the determination of hyperon and neutron beta decays~\cite{Ratcliffe:1998su}:
$D=0.80$ and $F=0.46$.  For more details on the notation and derivation of these chiral Lagrangians the reader is referred to \cite{Oller:2006yh}.

The perturbative amplitudes $V_{ij}$ for the meson-baryon scattering $(\phi B)_i \to (\phi B)_j$ up to $\cO(p^2)$  have been given in Refs.~\cite{Oller:2000fj,Oller:2006jw,Borasoy:2005ie}. Here the labels $i,j=1,2,3,\ldots,10$ correspond to the ten different channels $\pi^0 \Lambda$ (1), $\pi^0 \Sigma^0$ (2), $\pi^- \Sigma^+$ 
(3), $\pi^+ \Sigma^-$ (4), $K^- p$ (5), $\bar{K}^0 n$ (6), $\eta \Lambda$ (7), $\eta \Sigma^0$ (8), $K^0 \Xi^0$ (9) and 
$K^+ \Xi^-$ (10). 
Taking the initial center of mass (CM)  three-momentum along the $z$-axis, the $S$-wave amplitudes $\mathcal{T}_{ij}(W)$ can be  projected out through
\begin{eqnarray}\label{defpwt}
\mathcal{T}_{ij}(W)=\frac{1}{4\pi} \int d\Omega \,V_{ij}(W,\Omega,\sigma,\sigma)\,,
\end{eqnarray}
where $W$ is the energy in CM, $\sigma$ is the third component of the spin 
for the initial and final baryons and $\Omega$ is the solid angle of the scattered final three-momentum. 
Notice that for an $S$-wave amplitude, the third component of the spin of the initial and final baryons is the same in Eq.~\eqref{defpwt}.  
We point out that in the following discussions the $S$-wave amplitudes  $(\phi B)_i \to (\phi B)_j$ will be exploited to study the corresponding cross sections.  
This should be justified in our case since we only consider data near the threshold of the reaction.

In $SU(3)$ baryon ChPT the resummation of the unitarity chiral loops is crucial because of the large 
mass of the $s$ quark and the corresponding larger masses of those pseudoscalar mesons and baryons with strangeness.
Then, in many kinetic configurations the masses of the two particles in a given channel are much larger than 
the typical three-momentum and this enhances the two-particle reducible 
loop contributions \cite{weinberg91npb}. In addition, we are 
also interested in the chiral dynamics involving the resonance region where the unitarity upper bound in partial waves could be saturated. Thus, it is not appropriate to treat unitarity in a perturbative way 
as in plain ChPT. The presence of the $\Lambda(1405)$, which is near below  the $\bar{K}N$ threshold, clearly signals that 
the perturbative chiral amplitudes can not be appropriate here and a nonperturbative method must be developed 
to probe the meson-baryon dynamics in the strangeness $-1$ sector. As a result, one must resum the right-hand cut that stems from  unitarity and for that we employ Unitary ChPT (UChPT). This is based on an approximate algebraic solution to the N/D method~\cite{Oller:1998zr,Oller:1997ti,Oller:1999me}, which was first applied to meson-meson interactions and then  to meson-baryon scattering~\cite{Oller:2000fj,Oller:2005ig,Oller:2006jw,Oset:1997it}.

 The unitarized meson-baryon scattering amplitude in this formalism
can be cast in  matrix notation as \cite{Oller:2000fj}
\begin{eqnarray} \label{defut}
 T(W) = \big[ 1 + N(W)\cdot g(W^2) \big]^{-1}\cdot N(W)~.
\end{eqnarray}
The function  $g(s)$, $s=W^2$,  collects the unitarity cuts contributed by the two-particle intermediate states 
and $N(W)$ only contains the crossed-channel cuts. The unitarity loop $g(s)$ can be calculated through a once subtracted dispersion relation 
(as well as in the dimensional regularization replacing the divergence by a constant)
and for the $i_{th}$ channel its explicit form reads
 \begin{align}\label{gs}
16\pi^2 g(s)_i &=  a_i(\mu)+\log\frac{m_i^2}{\mu^2} 
-x_+\log\frac{x_+-1}{x_+}
-x_- \log\frac{x_--1}{x_-} \,, \\ 
x_{\pm} &=\frac{s+M_i^2-m_i^2}{2s}\pm\frac{1}{2s}\sqrt{ -4 s (M_i^2-i0^+)+(s+M_i^2-m_i^2)^2}\,, \nonumber 
\end{align}
where $a_i(\mu)$ is the subtraction constant that will be fitted to data, $M_i$ and $m_i$ denote the 
baryon and meson masses in the $i_{th}$ channel. We point out that $g(s)$ 
is independent of the scale $\mu$, introduced for dimensional reasons. Note that the combination $a_i(\mu)+\log m_i^2/\mu^2$ 
could be reabsorbed in a scale independent subtraction constant. In our following discussions, 
we set $\mu=770$~MeV, that fixes the scale at which the subtraction constants $a_i(\mu)$ are determined. 
We also constrain these subtraction constants by taking into account that when isospin symmetry is conserved the different $a_i$ attached to the states with different charges but 
made up from the same type of pseudoscalars and baryons should be the same.  This was demonstrated for $SU(3)$ symmetry in Ref.~\cite{Jido:2003cb} 
and a similar proof can be given straightforwardly for the isospin $SU(2)$ symmetry, a subgroup of $SU(3)$. 
In this way, the three subtraction constants for $\pi^0\Sigma^0$, $\pi^-\Sigma^+$ and $\pi^+\Sigma^-$  
are the same and equal to $a_2$. The two subtraction constants for $K^-p$ and $\bar{K}^0n$ are denoted by $a_5$ and 
those for $K^0\Xi^0$ and $K^+\Xi^-$ correspond to $a_9$.

Concerning $N(W)$ in Eq.~\eqref{defut}, it does not contain any unitarity cut and in our present discussion 
it is simply equal to the perturbative partial wave amplitude ${\cal T} (W)$ in Eq.~\eqref{defpwt} calculated up to ${\cal O}(p^2)$. However, for a higher order calculation of the input chiral amplitude ${\cal T}(W)$, the interaction kernel $N(W)$ is different from ${\cal T}(W)$ and the corresponding formula relating both can 
 be found in Ref.~\cite{Oller:2000fj}.

With the previous set up, we are ready to calculate the cross section  $(\phi B)_i\to (\phi B)_j$, which in our normalization reads
\begin{equation}\label{defcrosssection}
 \sigma[(\phi B)_i\to (\phi B)_j] = \frac{1}{16\pi\,s} \frac{|\vec{p_j}|}{|\vec{p_i}|} |T_{(\phi B)_i\to (\phi B)_j}|^2\,.
\end{equation}
 In the previous equation $T_{(\phi B)_i\to (\phi B)_j}$ corresponds to the unitarized $S$-wave amplitude in Eq.~\eqref{defut},   
$\vec{p_i}$ and $\vec{p_j}$  denote the initial and final CM three-momentum of the baryons, respectively.  We fit the cross sections of eight different processes: 
$\sigma(K^- p \to K^- p)$, $\sigma(K^- p \to \bar{K}^0 n)$, $\sigma(K^- p \to \pi^+ \Sigma^-)$, $\sigma(K^- p \to \pi^- \Sigma^+)$, 
$\sigma(K^- p \to \pi^0 \Sigma^0)$, $\sigma(K^- p \to \pi^0 \Lambda)$, $\sigma(K^- p \to \eta \Lambda)$ 
and $\sigma(K^- p \to \pi^0\pi^0\Sigma^0)$. 

 In this work, we only consider the total cross sections near the energy region above the threshold. 
Near the threshold, the leading behavior of the partial wave amplitude ${T}^{l}(W)$, with the orbital angular momentum number $l$, 
is proportional to $|p|^{l}|p'|^l$, being $\vec{p}$ and $\vec{p}\,'$ the initial and final three-momenta of the baryons in CM. 
 For the higher three-momenta shown for the first 6 panels in Figs.~\ref{fig1}--\ref{figopcs} the geometric mean $\sqrt{p p'}$ is 
around 200~MeV, which is significantly smaller than the upper limit of the laboratory $K^-p$ three-momentum of 300~MeV.  
Note also that, a posteriori, our assumption of $S$-wave dominance is able to provide a 
good description of data in the higher energy region shown in the plots. 
This is achieved in a natural way without forcing the free parameters of the fit. 
This assumption that we take here is shared by any other previous study of $\bar{K}N$ scattering in coupled 
channels based on ChPT \cite{kaiser:1995,Oset:1997it,Oller:2000fj,Oller:2005ig,Oller:2006jw,GarciaRecio:2002td,Jido:2003cb,Borasoy:2004kk,Borasoy:2005ie,Magas:2005vu,Hyodo:2002pk,Ikeda:2011pi,Ikeda:2012au,Cieply:2011nq,Mai:2012dt}.
 Nevertheless, if one attempts to study the differential cross sections, 
the higher partial waves become crucial, since only higher partial waves encompass the nontrivial angular distribution information and, furthermore, they add coherently so that larger interference effects are present.

Two event distributions are also fitted: the $\pi^-\Sigma^+$ event distribution 
from $K^- p \to \Sigma^+(1660)\pi^- \to \Sigma^+\pi^-\pi^+\pi^-$ \cite{hemmingway85npb} 
and the $\pi^0\Sigma^0$ event distribution from $K^- p \to \pi^0\pi^0\Sigma^0$ \cite{prakhov04prc}. 
For the  $K^- p \to \pi^0\pi^0\Sigma^0$ reaction we consider the proton pole dominant exchange  mechanism developed in Ref.~\cite{Magas:2005vu}, and evaluated as in Ref.~\cite{Oller:2006jw} (see Fig.~3 of the latter reference ).\footnote{In this exchange the proton is almost on-shell and this is the reason why the proton pole exchange mechanism is expected to be dominant \cite{Magas:2005vu}.}
We normalize the $\pi^0\Sigma^0$ event distribution by multiplying  our theoretical results by a constant fixed in order to reproduce 
the highest peak  value in the event distribution. 

To describe the $\pi^-\Sigma^+$ event distribution in the $\Lambda(1405)$ energy region, we follow the scheme developed in Ref.~\cite{Oller:2000fj} based on the assumption that the process takes places from an original scalar isoscalar source (with the same quantum numbers as the $\Lambda(1405)$) which is then corrected by 
final state interactions involving the three $\pi\Sigma$ and the two $\bar{K}N$ channels. For more details we refer to
 Ref.~\cite{Oller:2000fj}. Two extra free parameters $r$ and $r'$ are then introduced and the resulting 
formula is \cite{Oller:2000fj} 
\begin{equation}
\frac{dN_{\pi\Sigma}}{dW}=\bigg|r(D^{-1}|_{32}+D^{-1}|_{33}+D^{-1}|_{34})+r'(D^{-1}|_{35}+D^{-1}|_{36})\bigg|^2 |\vec{p}_{\pi^-\Sigma^+}|~.
\label{ompis} 
\end{equation}
In this equation we indicate by $D^{-1}|_{ij}$ the matrix elements of the inverse of the matrix 
$D=[1 + \mathcal{T}(W)\cdot g(W^2)]$.\footnote{Notice that the $T$-matrix in Eq.~\eqref{defut} can be written as 
$T(W)=D(W)^{-1}N(W)$.} 

Three ratios between different cross sections at the $K^- p$ threshold are also included in the fit \cite{tovee71npb,nowak78npb}: 
\begin{eqnarray}
\gamma&=&\frac{\sigma(K^-p\rightarrow \pi^+\Sigma^-)}
{\sigma(K^-p\rightarrow \pi^-\Sigma^+)}~, \quad
R_c=\frac{\sigma(K^-p\rightarrow \hbox{charged particles})}
{\sigma(K^-p\rightarrow \hbox{all})},~\quad
R_n=\frac{\sigma(K^-p\rightarrow \pi^0\Lambda)}
{\sigma(K^-p\rightarrow \hbox{all neutral states})}~.
\label{ratios}
\end{eqnarray}

The formulas for $\sigma_{\pi N}$, $a^{+}_{0+}$ and the masses of $N$, $\Lambda$, $\Sigma$ and $\Xi$ 
at $\cO(p^2)$ level calculated from the Lagrangians ${\cal L}_1$ and ${\cal L}_2$ in Eq.~\eqref{lagp2}, read 
\begin{eqnarray}
\sigma_{\pi N}&=&-2 M_\pi^2(2b_0+b_D+b_F)\,,\nn\\
a_{0+}^+&=&-\frac{M_\pi^2}{2\pi f^2}\left[(2b_0+b_D+b_F)-(b_1+b_2+b_3+2 b_4)+\frac{(D+F)^2}{8 m_p}\right]\,,\nn\\
m_N&=&m_0-2(b_0+2b_F)M_\pi^2-4( b_0+ b_D-b_F)M_K^2\,,\nn\\
m_{\Lambda}&=&m_0-\frac{2}{3}(3b_0-2b_D)M_\pi^2-\frac{4}{3}(3b_0+4b_D)M_K^2\,,\nn\\
m_{\Sigma}&=&m_0-2(b_0+2b_D)M_\pi^2 - 4b_0 M_K^2\,,\nn\\
m_{\Xi}&=&m_0-2(b_0-2b_F)M_\pi^2-4( b_0+ b_D+b_F)M_K^2\,,
\label{opd}
\end{eqnarray}
where we do not consider isospin-breaking effects for these quantities. The expressions for the masses of the baryons depend on the baryon mass in the $SU(3)$ 
chiral limit denoted by $m_0$, which always enter as $m_0-2b_0(m_\pi^2+2m_K^2)$. Thus, from the study of masses one can not disentangle both parameters $m_0$ and $b_0$, although  the latter
can be fitted from scattering. In our study we fix $m_0=0.9\pm 0.2$~GeV from the higher order study \cite{Frink:2005ru}, as in Ref.~\cite{Oller:2006jw}. Its uncertainty is propagated to the errors in the final results  obtained. Let us note that for the fits employing the interacting kernel calculated at ${\cal O}(p)$ all the next-to-leading (NLO) counterterms $b_0$, $b_D$, $b_F$ and $b_i$, $i=1,2,3,4$, are set to zero in Eq.~\eqref{opd} and none of these quantities are then introduced in the fits at ${\cal O}(p)$ described below.

Finally, the new SIDDHARTA measurement on the energy shift $\Delta E$ and width $\Gamma$ of 
the   kaonic hydrogen $1s$ state are related to the $K^- p$ 
scattering length $a_{K^- p}$, after taking into account isospin-breaking corrections~\cite{Meissner:2004jr}, through
\begin{eqnarray}\label{defegamma}
 \Delta E - i\frac{\Gamma}{2} = -2\alpha^3 \,\mu_r^2 \,a_{K^- p} \big[ 1 + 2 \alpha \,\mu_r \,(1-\ln\alpha) \,a_{K^- p} \big]\,,
\end{eqnarray}
where $\alpha$ is the fine structure constant and $\mu_r= m_p M_{K^-}/(M_{K^-}+ m_p)$ is the reduced mass of the $K^-$ and $p$ system. 
 In our normalization the relation between the $K^- p$ scattering length and the unitarized 
amplitude Eq.~\eqref{defut} reads  
\begin{equation}\label{defscattl}
 a_{K^- p} = \mathcal{F} (\sqrt{s}) \equiv \frac{T_{K^-p\to K^-p}(\sqrt{s})}{8\pi \sqrt{s}}\,, \quad \sqrt{s}=M_{K^-} + m_p\,.
\end{equation}

\section{Fits and results}
\label{fitresult}

Once we have introduced the theoretical formalism and the data included in the fits, we now proceed
 with the phenomenological discussions. 
In order to fix the free parameters we consider two different strategies to perform the fits, taking into account the fact 
that in literature there are two common ways to treat the light meson decay constants. For example, 
only one single decay constant is used for $\pi$, $K$ and $\eta$ in 
Refs ~\cite{kaiser:1995,Oset:1997it,GarciaRecio:2002td,Oller:2000fj,Oller:2006jw,Magas:2005vu,Jido:2003cb}, 
while physical values are adopted in Refs.~\cite{Nieves:2001wt,Ikeda:2011pi,Ikeda:2012au,Mai:2012dt}. 
In Fit I, we use a common decay constant for all the states  in 
the meson-baryon scattering amplitudes  and  then fit this single decay constant. 
In Fit II, we distinguish the decay constants for 
$\pi$, $K$ and $\eta$, according to the pseudoscalar appearing in the state, so that one then employs $f_\pi$, $f_K$ and $f_\eta$, in order. 
The values $f_\pi=92.2$~MeV and $f_K=110.0$~MeV are taken from the PDG~\cite{pdg}, while  
$f_\eta=120$~MeV (in fact it is $f_{\eta_8}$) is estimated by using the result from Ref.~\cite{Kaiser:1998ds}. 
Note that both strategies are compatible with the calculation of ${\cal T}$ in ChPT up to ${\cal O}(p^2)$,
 since the differences induced by using one or other decay constants are at least of ${\cal O}(p^3)$. 
For the masses of mesons and baryons, we always take their physical values in the scattering amplitudes 
in order to properly account for the threshold effects.

To equally weight experimental data from different measurements, we divide the $\chi^2$ from 
each measurement by its number of data points and then sum over all the partial $\chi^2$ together. 
Then, the $\chi^2$ per degree of freedom ($\chi^2_{d.o.f}$) is defined as 
\begin{align}\label{defchisq}
\chi^2_{d.o.f} &= \frac{\sum_k n_k}{K\,(\sum_k n_k - n_p)}\sum_{k=1}^K \frac{\chi^2_k}{n_k}~,\nn\\
\chi^2_k&=\sum_{i=1}^{n_k} \frac{\left(y_{k;i}^{th}-y_{k;i}^{exp}\right)^2}{\sigma_{k;i}^2}~.
\end{align}
Here $K$ is the number of different measurements considered in our fits, $n_k$ stands for 
the number of data points in the $k_{\rm th}$ measurement,  $n_p$ corresponds to 
the number of free parameters and $y_{k;i}^{exp}\,(y_{k;i}^{th})$ represents the $i_{\rm{th}}$ experimental (theoretical) point of the $k_{\rm{th}}$ set of data with standard deviation $\sigma_{k;i}$. This definition of $\chi^2$ is also used 
in Refs.~\cite{GarciaRecio:2002td, Borasoy:2005ie,Ikeda:2011pi,Ikeda:2012au,Mai:2012dt}. 
 This is basically done in order to enhance the weight of the kaon hydrogen data in the fits performed, because one 
assumes that this is a high quality data point. Nonetheless, we consider that using the previous 
 definition of $\chi^2_{d.o.f}$ in Eq.~\eqref{defchisq} instead of  the standard one for a total set 
of $\sum_{k=1}^K n_K$ data points,
\begin{align}
\chi^2_{d.o.f}=\frac{1}{\sum_k n_k-n_p}\sum_{k=1}^K \chi^2_k~,
\label{defchisq2}
\end{align}
is somewhat arbitrary. In this way, we will also discuss the stability of the fits by switching from Eq.~\eqref{defchisq} to Eq.~\eqref{defchisq2}.

Concerning the experimental data, the references for the cross sections 
and event distributions are explained in Fig~\ref{fig1}, while the remaining were already introduced above. For the ratio $\gamma$ measured 
at the $K^-p$ threshold in Eq.~\eqref{ratios}, we notice that Ref.~\cite{tovee71npb} 
reported the value $2.34\pm0.08$ and Ref.~\cite{nowak78npb} gave two values:~$2.38\pm0.04$ and $2.35\pm0.07$. 
Other value  $\gamma=2.15$ can be also found in Ref.~\cite{humphrey62pr}. 
Hence in our current work we decide to assign $\gamma$ a conservative error band at $5\%$ level, 
i.e. $\gamma=2.36\pm0.12$, instead of $\gamma=2.36\pm0.04$ used in Refs.~\cite{Borasoy:2005ie,Oller:2006jw,Cieply:2011nq,Ikeda:2012au} 
and $\gamma=2.38\pm0.04$ in Ref.~\cite{Mai:2012dt}.  
A similar conservative error bar is also assigned for $R_c$, namely $R_c=0.664\pm0.033$.   
While since the figure $R_n=0.189\pm0.015$ derived in Ref.~\cite{martin81npb} already contains an error bar larger 
than our conservative $5\%$ relative error. In connection to this, we do not consider that we could have phenomenology well 
under control within a precision of 5\% or better.  
As mentioned previously, in order to account for our theoretical uncertainties, we assign large errors for the pion 
nucleon isospin even $S$-wave scattering length $a_{0+}^{+}$, nucleon sigma term $\sigma_{\pi N}$, and 
the masses of $N, \Lambda, \Sigma$ and $\Xi$, whose central values and ascribed error bars can be found in the second column of Table \ref{tabresult}.  
For $\sigma_{\pi N}$ one expects that it receives sizable higher-order corrections from the mesonic cloud which are expected to be positive and around 10~MeV \cite{leu_sainio}. On the other hand, the precise value for this observable is still under debate and one has the classic result $\sigma_{\pi N}=45\pm 8$~MeV \cite{leu_sainio} and the one favored 
from more modern experimental $\pi N$ data bases $\sigma_{\pi N}=59\pm 7$~MeV \cite{alarcon:2012}. This adds another 10~MeV of uncertainty, so that in our fits employing an ${\cal O}(p^2)$ calculation for this quantity we take $\sigma_{\pi N}=30\pm 20$~MeV. By including such a large errorbar the central value for $\sigma_{\pi N}$ is not really relevant and its inclusion in our study is useful only to discard fits that would give rise to really awful $\sigma_{\pi N}$. Regarding the isoscalar scalar $\pi N$ scattering length its latest recent determination from pionic atoms is $a_{0+}^+=(7.6\pm 3.1)\cdot 10^{-3}$~$m_\pi^{-1}$ \cite{baru:2011}. Similarly as in Ref.~\cite{Oller:2006jw} we expect a theoretical error in our ${\cal O}(p^2)$ calculation of around $+1\cdot 10^{-2}$~$m_\pi^{-1}$, estimated from the ${\cal O}(p^3)$ unitarity corrections \cite{Bernard:1993}. As a result we take in our fits the value $a_{0+}^+=(0\pm 1)\cdot 10^{-2}$~$m_\pi^{-1}$. For the masses we include an error of 30\% by considering 
a general expectation for the breaking of an $SU(3)$ prediction.

Now we are ready to present our fit results. The error bars of the parameters in the fits  represent only the statistical uncertainty 
at the level of one standard deviation. For the $\chi^2$ definition Eq.~\eqref{defchisq2}, with a large numbers of d.o.f, we employ the criteria 
\begin{align}
\chi^2 \leq \chi_0^2 + n_\sigma\sqrt{2\chi_0^2}\,,
\label{nsigma}
\end{align} 
to calculate the parameter intervals. 
In the previous equation,  $n_\sigma$ is the number of standard deviations, 
$\chi_0^2$ is the minimum $\chi^2$ calculated with the central values of the fit 
and $\chi^2$ results by  taking a new configuration of free parameters. 
 Eq.~\eqref{nsigma} was deduced in the appendix of Ref.~\cite{etkin}, making use of the fact that the quantity $(\chi^2-\nu)/(\sqrt{2\nu})$ 
is normally distributed with the mean value zero and standard deviation one in the limit of a large number $\nu$ of d.o.f, i.e. the number 
of data points minus the number of free parameters.  For a good fit, the minimum chi-square $\chi_0^2$ should approach to the number of d.o.f $\nu$.

For the definition of the $\chi^2$ in Eq.~\eqref{defchisq}, we note that the effective number of data points is quite small and the criteria 
given in Eq.~\eqref{nsigma} may not be appropriate. Therefore, we take another strategy  for the estimation of the parameter 
intervals at a given confident level, 
which was originally developed in the astrophysical analysis~\cite{avni}.  
The basic idea is that $\Delta {\chi^2} \equiv \chi^2 - \chi_0^2$, obeys the chi-square distribution with $n_p$ d.o.f, being $n_p$ the number of 
free parameters, because $\chi^2$ obeys a chi-square distribution with $N$ d.o.f, being $N$ the number of data points and the minimum 
chi-square $\chi_0^2$ from a fit is distributed with $N-n_p$ d.o.f. Detailed discussions can be found in Ref.~\cite{avni}. So in summary, for 
the $\chi^2$ definition in Eq.~\eqref{defchisq}, we use the criteria 
\begin{align}
\chi^2 \leq \chi_0^2 + \Delta\chi^2\,,
\label{deltachisq}
\end{align} 
to estimate the parameter intervals, being $\Delta\chi^2$ a chi-square distribution with $n_p$ d.o.f.

We point out that in both cases we calculated the correlated error bars, instead of estimating the uncertainty of a single parameter one by one, since 
we generate new parameter configurations by randomly varying all the free parameters around their central values through a Monte Carlo generator. 
With these new configurations, we then calculate the new values of $\chi^2$ and reject those configurations with a $\chi^2$ larger than the upper 
limit given above, corresponding to a confidence level of one sigma. In the particular 
case of Eq.~\eqref{nsigma} one has to place $n_\sigma=1$ and discount those fits with $\chi^2 >\chi^2_0+\sqrt{2\chi_0^2}$. The  set of 
parameter configurations with lower $\chi^2$ will be kept and used  in later discussions for determining the uncertainty in any observable considered. 
In such a way, our estimation of the error bars implicitly takes into account the correlated errors of all the free parameters.

\subsection{Results using the $\chi^2$ defined  in Eq.~\eqref{defchisq}}

We first discuss the results 
employing the definition for the $\chi^2$ corresponding to Eq.~\eqref{defchisq}, in  which every set of data corresponding to 
one observable is considered as an  effective data point. The fitted free parameters are given in Table~\ref{tabpara} for the different fits considered, namely, for the leading order fit, ${\cal O}(p)$-Fit, and for the NLO ones,  Fit I and Fit II. From this table we observe that the $b_i$ coefficients have natural values which are quite similar between the 
two fits I and II (of course for the ${\cal O}(p)$-Fit they are fixed to zero).  The largest differences happen for the values of the subtraction constants, 
specially for $a_1$. Nonetheless, one has to stress that all the subtraction constants in our fits have natural size of ${\cal O}(1)$.
  The cross sections and event distributions are shown by the (red) solid 
lines in Figs.~\ref{fig1}, \ref{figfit2cs} and \ref{figopcs} for the fits I, II and ${\cal O}(p)$-Fit, in order. The hatched area around each line is the estimated statistical uncertainty at the level of one sigma as explained above.  In Table \ref{tabresult} we give the reproduction for the rest of observables fitted, and show separately the energy shift and width of the $1s$ state of kaonic hydrogen state in the left panel of Fig.~\ref{fig:squares} by the empty symbols, as indicated in the figure.  The rest of contents in Table~\ref{tabresult} are shown by the empty squares and circles in Fig.~\ref{figratios} for fits I and II, respectively. In this figure  every observable is appropriately rescaled so that all of them have similar size and the quality of its reproduction can be easily appreciated. 

As it is clear from Figs.~\ref{fig1}, \ref{figfit2cs} and Table~\ref{tabresult}  the reproduction of data for fits I and II is very good with a $\chi^2_{d.o.f}$  lower than 1, see the first line in Table~\ref{tabpara}. The reproduction of the observables is then consistent between each other. In this respect we find that our results are compatible
 with the value of $\delta_{\pi \Lambda}$ from Ref.~\cite{e756} and not with Ref.~\cite{huang}.  It is also remarkable that the ${\cal O}(p)$-Fit, having 6 free parameters less than Fit II and 7 less than Fit I, is able to achieve a good reproduction of data, with the important exception of  the $K^-p\to \eta\Lambda$ cross section that  it is not able to reproduce adequately. Indeed, this is the reason why the $\chi^2_{d.o.f}$ for the ${\cal O}(p)$-Fit is near 2. If the data of the cross section for $K^-p\to\eta\Lambda$ is removed the resulting $\chi^2_{d.o.f}$ is 1.23, a much lower value. It follows then that the ${\cal O}(p)$-Fit  can properly reproduce the lower-energy scattering data as well as the SIDDHARTA measurement on the energy shift and width of the $1s$ state of kaonic hydrogen. 
  As a result, one finds an indication towards convergence in the reproduction of all these data since it can already be well reproduced by the ${\cal O}(p)$-Fit (except the $K^-p\to \eta\Lambda$ cross section) and the description is improved when including ${\cal O}(p^2)$ contributions to the interacting kernel (in which case the $K^-p\to \eta\Lambda$ can also be accounted for). This indication of convergence in the reproduction of data when passing from the leading to NLO fits did not result when fitting the DEAR data \cite{dear} on the energy width and shift of the $1s$ state of kaonic hydrogen \cite{Oller:2006jw}. We also find that the  
ratios in Eq.~\eqref{ratios} are improved in Table~\ref{tabresult} compared to the results 
obtained in Ref.~\cite{Oller:2006jw}. 
The masses of the lightest baryon octet and $\sigma_{\pi N}$ are satisfactorily reproduced as well.  
Our results for $\Delta E$ and $\Gamma$ in Eq.~\eqref{defegamma} are compatible with the SIDDHARTA 
measurements within errors, though our fits prefer somewhat lager central values for $\Gamma$, 
as shown in Table~\ref{tabresult} and Fig.~\ref{fig:squares}. It is clear from the right panel of Fig.~\ref{fig:squares} that the results from other  recent studies \cite{Ikeda:2011pi,Ikeda:2012au,Cieply:2011nq,Mai:2012dt} also prefer larger values for $\Gamma$.

We also calculate the $K^-p$ scattering length $a_{K^- p}$, Eq.~\eqref{defscattl}, and 
those with $I=0$ and $I=1$ extracted from data and the theoretical model. 
The results are shown in Table \ref{table:scat}. Our values are compatible with the recent 
determinations from Refs.~\cite{Ikeda:2012au,Mai:2012dt}. In addition, we show in Fig.~\ref{fig2} the 
 $K^-p\to K^-p$ $S$-wave amplitude around the threshold region and below it, the left panel is for the real part and the right one for the imaginary part.  The points with error bars in Fig.~\ref{fig2} correspond 
to $a_{K^-p}$ extracted directly from SIDDHARTA data by making use of Eq.~\eqref{defegamma}, $a_{K^-p}=(-0.65\pm 0.10)+i\,(0.81\pm 0.15)$~fm.   We observe that this amplitude for fits I and II is quite the same at threshold and above it in the range shown in the figure, indeed 
 the threshold parameters from both fits perfectly agree with each other in Table \ref{table:scat}. 
 However, quite different behaviors result in the energy region below threshold.  
The real parts of the $K^-p\to K^-p$ scattering amplitudes from the two fits become incompatible below  
 1.42~GeV and the imaginary parts are also incompatible below 1.41~GeV. As a result, it is clear that a precise knowledge of the $a_{K^-p}$ scattering length, 
as that resulting from the SIDDHARTA data, can not pin down in a precise manner the extrapolation of the $K^-p$ $S$-wave amplitude below threshold. 
Indeed, the difference from Fit I and II for the subthreshold extrapolation of the $K^-p$ $S$-wave amplitude in Fig.~\ref{fig2} is much larger than the uncertainty 
estimated in Refs.~\cite{Ikeda:2011pi,Ikeda:2012au,Cieply:2011nq,Mai:2012dt}.  However, an ${\cal O}(p^3)$ determination of the interacting kernel $N(W)$ in UChPT, Eq.~\eqref{defut},  could reduce this uncertainty because 
then one is sensitive to the change of the weak decay constants so that these changes have to be compensated by new terms at ${\cal O}(p^3)$. This is certainly an 
interesting calculation in order to improve the accuracy of our knowledge in this field of research.

\begin{table}[H]
\renewcommand{\tabcolsep}{0.5cm}
\renewcommand{\arraystretch}{1.3}
{\footnotesize
\begin{center}
\begin{tabular}{|c|c|c|c|}
\hline 
Parameters & Fit I & Fit II &  $ \cO(p)$-Fit
\\
  & $\chi^2_{d.o.f}=0.85$ & $\chi^2_{d.o.f}=0.96$  & $\chi^2_{d.o.f}=1.87$
\\
\hline 
$f$ (MeV)           & 124.60$_{-1.58}^{+1.84}$  &  Fixed  &  116.05$_{-1.59}^{+1.89}$
\\
\hline 
$b_0$ (GeV$^{-1}$)  & -0.230$_{-0.026}^{+0.029}$  &  -0.292$_{-0.007}^{+0.008}$ &  0
\\ 
\hline 
$b_D$ (GeV$^{-1}$)  & -0.027$_{-0.020}^{+0.025}$  &  0.101$_{-0.008}^{+0.010}$  &  0
\\ 
\hline 
$b_F$ (GeV$^{-1}$)  & -0.183$_{-0.082}^{+0.094}$ &  -0.200$_{-0.011}^{+0.011}$  &  0
\\ 
\hline 
$b_1$ (GeV$^{-1}$)  & 0.714$_{-0.019}^{+0.011}$  &  0.522$_{-0.006}^{+0.005}$   &  0
\\  
\hline 
$b_2$ (GeV$^{-1}$)  & 1.331$_{-0.048}^{+0.051}$  &  1.015$_{-0.023}^{+0.024}$   &  0
\\  
\hline 
$b_3$ (GeV$^{-1}$)  & -0.696$_{-0.078}^{+0.078}$  &  -0.306$_{-0.014}^{+0.015}$ &  0
\\[0.1cm] 
\hline 
$b_4$ (GeV$^{-1}$)  & -0.889$_{-0.039}^{+0.037}$  &  -0.899$_{-0.012}^{+0.010}$ &  0
\\ 
\hline 
$a_1$  & 2.587$_{-0.944}^{+0.962}$   &  4.761$_{-0.397}^{+0.491}$   &  -6.377$_{-1.288}^{+1.509}$
\\ 
\hline 
$a_2$  & -0.830$_{-0.140}^{+0.134}$  &  -0.447$_{-0.163}^{+0.168}$  &  -1.772$_{-0.186}^{+0.236}$
\\ 
\hline 
$a_5$  & -1.073$_{-0.030}^{+0.034}$  &  -1.685$_{-0.041}^{+0.042}$  &  -1.668$_{-0.042}^{+0.048}$
\\  
\hline 
$a_7$  & 1.164$_{-0.355}^{+0.488}$   &  1.401$_{-0.155}^{+0.185}$   &  -2.215$_{-0.149}^{+0.180}$
\\ 
\hline 
$a_8$  & -1.938$_{-1.400}^{+0.831}$  &  -0.168$_{-0.050}^{+0.076}$  &  -0.170$_{-0.239}^{+0.241}$
\\ 
\hline 
$a_9$  & -2.161$_{-0.022}^{+0.035}$  &   -2.406$_{-0.027}^{+0.038}$ &  -2.223$_{-0.068}^{+0.101}$
\\   
\hline 
$r$ (GeV$^{-1}$)  & 24.28$_{-5.39}^{+4.78}$  &  18.27$_{-4.95}^{+3.18}$  &  11.20$_{-13.96}^{+4.14}$ 
\\ 
\hline 
$r'$ (GeV$^{-1}$) & 10.85$_{-6.72}^{+6.59}$  &  17.65$_{-14.06}^{+12.26}$  &  5.40$_{-18.94}^{+6.18}$
\\
\hline 
\end{tabular}
 \caption{ Parameters from the three fits with the $\chi^2$ defined in Eq.\eqref{defchisq}. 
The way in which the error bars of parameters are calculated is  explained in the text. }
\label{tabpara}
\end{center}
}
\end{table}

\begin{table}[H]
\renewcommand{\tabcolsep}{0.07cm}
\renewcommand{\arraystretch}{1.3}
{\footnotesize
\begin{center}
\begin{tabular}{|c|c|c|c|c|}
\hline 
Observable & Input & Fit I & Fit II &  $\cO(p)$-Fit
\\
\hline 
$\Delta E$ (eV) & 283$\pm$36 & 299$_{-41}^{+33}$  &  276$_{-46}^{+48}$   &  276$_{-46}^{+55}$
\\  
\hline 
$\Gamma$ (eV) &  541$\pm$92  & 612$_{-66}^{+66}$  &  608$_{-65}^{+50}$   &  606$_{-72}^{+72}$
\\ 
\hline 
$\gamma$ & 2.36$\pm$0.12  & 2.36$_{-0.22}^{+0.17}$  &  2.36$_{-0.23}^{+0.24}$    &  2.36$_{-0.28}^{+0.27}$
\\
\hline 
$R_c$ & 0.664$\pm$0.033 & 0.662$_{-0.010}^{+0.008}$  &  0.661$_{-0.011}^{+0.012}$    &  0.647$_{-0.007}^{+0.006}$
\\ 
\hline 
$R_n$ & 0.189$\pm$0.015 & 0.192$_{-0.020}^{+0.025}$  &  0.188$_{-0.029}^{+0.028}$    &  0.188$_{-0.032}^{+0.033}$
\\ 
\hline 
$\delta_{\pi \Lambda}$ (degrees) & $3.2\pm 5.3$ &  -1.1$_{-1.7}^{+1.1}$  & -2.7$_{-0.4}^{+0.5}$  &  -1.8$_{-0.4}^{+0.3}$ 
 \\
\hline
$a^{+}_{0+}$ ($10^{-2}m_\pi^{-1}$) & $0\pm$1.0 & 0.08$_{-0.45}^{+0.23}$  &  -0.46$_{-0.22}^{+0.22}$  &  -0.66$_{-0.02}^{+0.02}$
\\  
\hline 
$\sigma_{\pi N}$ (MeV) & 30$\pm$20  & 25.6$_{-6.0}^{+4.7}$  &  26.0$_{-1.0}^{+1.0}$ &  0
\\  
\hline 
$M_N$ (GeV) &  0.94$\pm$0.28       & 1.00$_{-0.22}^{+0.23}$  &  0.92$_{-0.20}^{+0.20}$  &  $m_0$
\\[0.1cm]  
\hline 
$M_\Lambda$ (GeV) & 1.12$\pm$0.34  & 1.17$_{-0.21}^{+0.21}$  &  1.07$_{-0.20}^{+0.20}$ &   $m_0$
\\ 
\hline 
$M_\Sigma$ (GeV) & 1.19$\pm$0.36   & 1.14$_{-0.20}^{+0.20}$ &  1.19$_{-0.20}^{+0.20}$  &   $m_0$
\\ 
\hline 
$M_\Xi$ (GeV) &  1.32$\pm$0.40     & 1.33$_{-0.24}^{+0.22}$  &  1.28$_{-0.20}^{+0.20}$   &  $m_0$
\\ 
\hline 
\end{tabular}
\caption{ Results by using the fits in Table~\ref{tabpara}. In the second column we give the values for the several observables as included in the fit.}
\label{tabresult}
\end{center}
}
\end{table}

\begin{table}[H]
\renewcommand{\tabcolsep}{0.2cm}
\renewcommand{\arraystretch}{1.3}
{\footnotesize 
\begin{center}
\begin{tabular}{|c|c|c|c|}
\hline
Observable & Fit I & Fit II & ${\cal O}(p)$-Fit \\
\hline
$a_{K^- p}$ (fm) & $-0.67_{-0.12}^{+0.13}+i\,0.92_{-0.08}^{+0.07}$  & $-0.61_{-0.15}^{+0.13}+i\,0.89_{-0.06}^{+0.07}$ & $-0.61_{-0.16}^{+0.15}+i\,0.89_{-0.07}^{+0.11}$ \\ 
\hline
$a_{I=0}$  (fm) & $-1.74_{-0.17}^{+0.20}+i\,1.27_{-0.12}^{+0.14}$  & $-1.58_{-0.31}^{+0.26}+i\,1.27_{-0.09}^{+0.09}$ & $-1.62_{-0.33}^{+0.31}+i\,1.32_{-0.12}^{+0.15}$ \\ 
\hline
$a_{I=1}$ (fm)  & $0.39_{-0.10}^{+0.12}+i\,0.56_{-0.09}^{+0.09}$   & $0.36_{-0.06}^{+0.06}+i\,0.52_{-0.09}^{+0.09}$ & $0.40_{-0.06}^{+0.05}+i\,0.45_{-0.08}^{+0.10}$ \\
\hline
\end{tabular}
\caption{
{\small $K^-p$ scattering length, $a_{K^-p}$, and $I=0$, $1$ $\bar{K}N$ scattering lengths, $a_{I=0}$, $a_{I=1}$, in order, for Fit I, Fit II and ${\cal O}(p)$-Fit given in Table~\ref{tabpara}.}
\label{table:scat}
} 
\end{center}
}
\end{table}

\begin{table}[H]
\renewcommand{\tabcolsep}{0.5cm}
\renewcommand{\arraystretch}{1.3}
{\footnotesize
\begin{center}
\begin{tabular}{|c|c|c|c|}
\hline 
Parameters & Fit I & Fit II &  $\cO(p)$-Fit
\\
  & $\chi^2_{d.o.f}=1.87$ & $\chi^2_{d.o.f}=1.93$  & $\chi^2_{d.o.f}=2.71$
\\
\hline 
$f$ (MeV)           & 125.71$_{-0.94}^{+1.25}$  &  Fixed  &  112.75$_{-2.14}^{+2.69}$
\\
\hline 
$b_0$ (GeV$^{-1}$)  & -0.169$_{-0.018}^{+0.019}$  &  -0.340$_{-0.009}^{+0.009}$ &  0
\\ 
\hline 
$b_D$ (GeV$^{-1}$)  & -0.108$_{-0.012}^{+0.016}$  &  0.196$_{-0.008}^{+0.011}$  &  0
\\ 
\hline 
$b_F$ (GeV$^{-1}$)  & -0.183$_{-0.051}^{+0.038}$ &  -0.265$_{-0.014}^{+0.010}$  &  0
\\ 
\hline 
$b_1$ (GeV$^{-1}$)  & 0.856$_{-0.006}^{+0.006}$  &  0.667$_{-0.007}^{+0.004}$   &  0
\\  
\hline 
$b_2$ (GeV$^{-1}$)  & 1.621$_{-0.027}^{+0.033}$  &  1.193$_{-0.017}^{+0.019}$   &  0
\\  
\hline 
$b_3$ (GeV$^{-1}$)  & -0.501$_{-0.049}^{+0.043}$  &  -0.504$_{-0.013}^{+0.013}$ &  0
\\[0.1cm] 
\hline 
$b_4$ (GeV$^{-1}$)  & -1.177$_{-0.027}^{+0.026}$  &  -1.082$_{-0.017}^{+0.012}$ &  0
\\ 
\hline 
$a_1$  & -0.800$_{-0.384}^{+0.435}$   &  3.241$_{-0.391}^{+0.391}$   &  -5.083$_{-0.596}^{+0.668}$
\\ 
\hline 
$a_2$  & -0.883$_{-0.086}^{+0.100}$  &  -0.351$_{-0.102}^{+0.103}$  &  -1.655$_{-0.135}^{+0.139}$
\\ 
\hline 
$a_5$  & -0.961$_{-0.045}^{+0.038}$  &  -1.558$_{-0.054}^{+0.071}$  &  -1.610$_{-0.074}^{+0.086}$
\\  
\hline 
$a_7$  & 0.651$_{-0.176}^{+0.217}$   &  1.275$_{-0.259}^{+0.213}$   &  -2.068$_{-0.195}^{+0.190}$
\\ 
\hline 
$a_8$  & -5.250$_{-2.807}^{+1.047}$  &  -0.168$_{-0.059}^{+0.064}$  &  -0.495$_{-0.221}^{+0.233}$
\\ 
\hline 
$a_9$  & -2.160$_{-0.013}^{+0.012}$  &   -2.298$_{-0.012}^{+0.014}$ &  -2.165$_{-0.060}^{+0.067}$
\\   
\hline 
$r$ (GeV$^{-1}$)  & 27.46$_{-3.60}^{+2.86}$  &  19.78$_{-2.12}^{+1.83}$ &  10.79$_{-5.16}^{+3.65}$ 
\\ 
\hline 
$r'$ (GeV$^{-1}$) & 11.84$_{-3.50}^{+1.77}$  &  21.76$_{-6.48}^{+6.32}$  &  7.02$_{-4.08}^{+3.89}$
\\
\hline 
\end{tabular}
 \caption{ Parameters from the three fits with the $\chi^2$ defined in Eq.\eqref{defchisq2}.  
 The error bars are calculated as explained in the text. }
\label{tabparacomchisq}
\end{center}
}
\end{table}

\begin{table}[ht]
\renewcommand{\tabcolsep}{0.2cm}
\renewcommand{\arraystretch}{1.3}
{\footnotesize
\begin{center}
\begin{tabular}{|c|c|c|c|c|}
\hline 
Observable & Input & Fit I & Fit II &  $\cO(p)$-Fit
\\
\hline 
$\Delta E$ (eV) & 283$\pm$36 & 334$_{-25}^{+21}$  & 324$_{-24}^{+28}$   &  308$_{-28}^{+36}$
\\  
\hline 
$\Gamma$ (eV) &  541$\pm$92  & 649$_{-50}^{+66}$  &  606$_{-43}^{+68}$   &  664$_{-79}^{+98}$
\\ 
\hline 
$\gamma$ & 2.36$\pm$0.12  & 2.21$_{-0.30}^{+0.40}$  &  2.17$_{-0.37}^{+0.37}$    &  2.00$_{-0.38}^{+0.47}$
\\
\hline 
$R_c$ & 0.664$\pm$0.033 & 0.652$_{-0.008}^{+0.007}$  &  0.649$_{-0.011}^{+0.009}$    &  0.640$_{-0.006}^{+0.007}$
\\ 
\hline 
$R_n$ & 0.189$\pm$0.015 & 0.227$_{-0.039}^{+0.038}$  &  0.198$_{-0.029}^{+0.041}$    &  0.207$_{-0.051}^{+0.052}$
\\ 
\hline 
$\delta_{\pi \Lambda}$ (degrees) & $3.2\pm 5.3$ &  -1.0$_{-1.1}^{+1.3}$  & -1.3$_{-0.4}^{+0.4}$  &  -1.7$_{-0.3}^{+0.3}$ 
 \\
\hline
$a^{+}_{0+}$ ($10^{-2}m_\pi^{-1}$) & $0\pm$1.0 &  0.10$_{-0.28}^{+0.21}$  &  -1.33$_{-0.16}^{+0.16}$  &  -0.70$_{-0.03}^{+0.03}$
\\  
\hline 
$\sigma_{\pi N}$ (MeV) & 30$\pm$20  & 23.9$_{-2.6}^{+2.8}$  &  28.5$_{-0.9}^{+1.0}$ &  0
\\  
\hline 
$M_N$ (GeV) &  0.94$\pm$0.28        & 1.01$_{-0.21}^{+0.21}$  &  0.81$_{-0.20}^{+0.20}$  &  $m_0$
\\[0.1cm]  
\hline 
$M_\Lambda$ (GeV) & 1.12$\pm$0.34  & 1.21$_{-0.20}^{+0.20}$  &  0.99$_{-0.20}^{+0.20}$  &   $m_0$
\\ 
\hline 
$M_\Sigma$ (GeV) & 1.19$\pm$0.36   & 1.08$_{-0.20}^{+0.20}$  &  1.23$_{-0.20}^{+0.20}$  &   $m_0$
\\ 
\hline 
$M_\Xi$ (GeV) &  1.32$\pm$0.40     & 1.34$_{-0.21}^{+0.21}$  &  1.29$_{-0.20}^{+0.20}$   &  $m_0$
\\ 
\hline 
\end{tabular}
\caption{   Results by using the fits in Table~\ref{tabparacomchisq}. In the second column we give the values for the several observables as included in the fit.
  }
\label{tabresultcomchisq}
\end{center}
}
\end{table}

\begin{figure}[H]
\begin{center}
 \includegraphics[angle=0, width=0.85\textwidth]{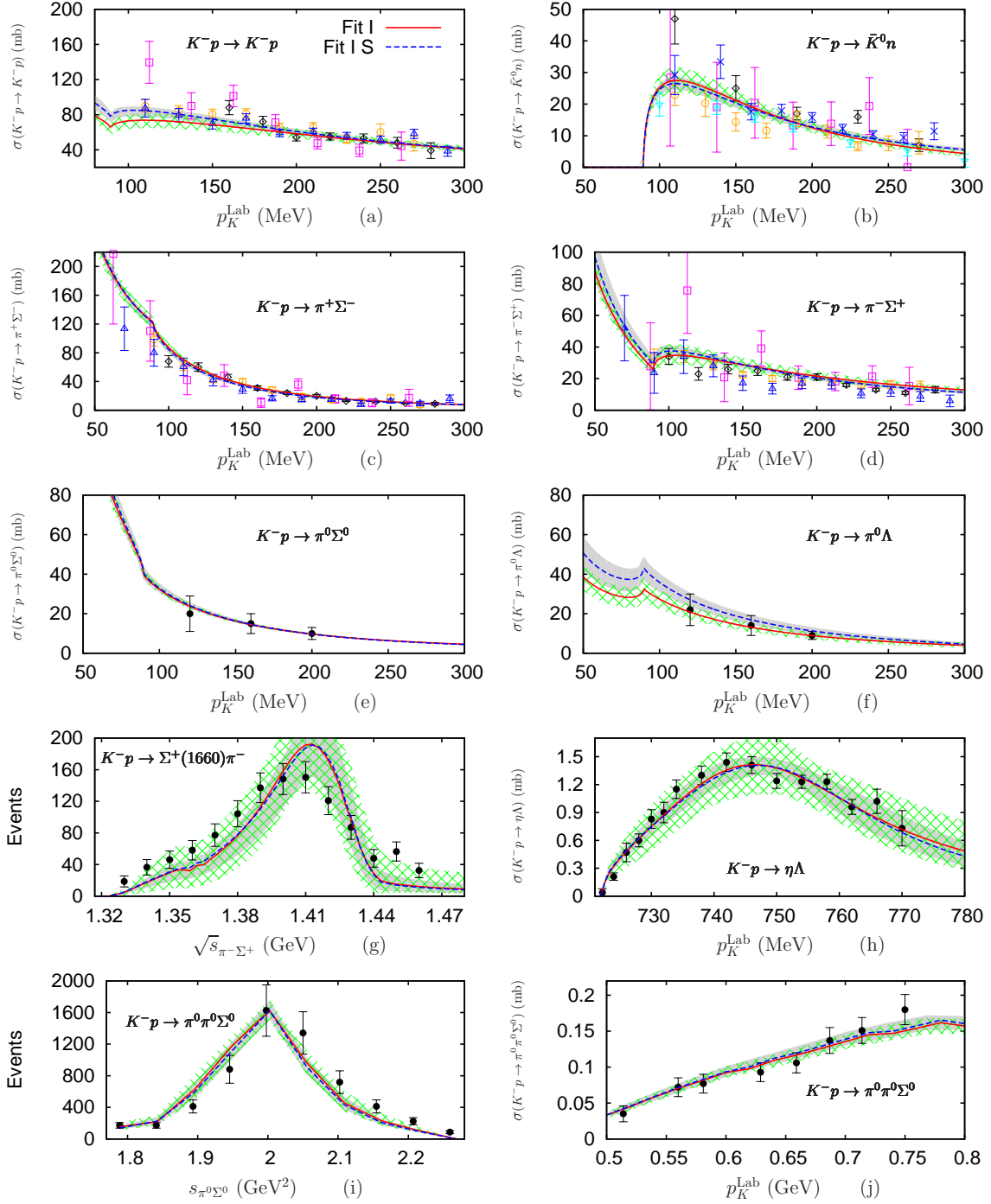}
\caption{{\small (Color online). The ten panels from (a) to (j) correspond to the cross sections of~$K^-p \to K^-p,\bar{K}^0n,$ 
$\pi^+\Sigma^-, \pi^-\Sigma^+, \pi^0\Sigma^0, \pi^0\Lambda$, 
the $\pi^-\Sigma^+$ event distribution from $K^-p \to \Sigma^+(1660)\pi^-$, the $K^-p \to \eta\Lambda$ cross section, 
the $\pi^0\Sigma^0$ event distribution from the reaction $K^-p\to\pi^0\pi^0\Sigma^0$ with $p_K=0.687$~GeV and the total cross section of 
$K^-p\to\pi^0\pi^0\Sigma^0$, respectively.  The data points represented by 
black diamond, magenta square, orange circle, blue cross, cyan down-triangle and blue up-triangle 
in the first four panels are taken from Refs.~\cite{ciborowsk82jpg,humphrey62pr,kim67prl,evans83jpg,kittel66pl,sakitt65pr},
respectively. The data in the fifth and sixth panels are from Ref.~\cite{kim66unirep}. 
The $\pi^-\Sigma^+$ event distribution is taken from Ref.~\cite{hemmingway85npb} and the 
$K^-p\to\eta\Lambda$ cross section data are given in Ref.~\cite{starostin01prc}. The measurements 
on the reaction $K^-p\to\pi^0\pi^0\Sigma^0$ are from Ref.~\cite{prakhov04prc}.  
The red solid lines and blue dashed lines represent the best fits from Fit I using Eqs.~\eqref{defchisq} and 
\eqref{defchisq2} (which is indicated by Fit~I~S), respectively. 
The areas covered by green hatched lines and the gray shaded areas correspond 
to our estimates of error bands for Fit I and Fit~I~S, in order.  }}
\label{fig1}
\end{center}
\end{figure}

\begin{figure}[H]
\begin{center}
 \includegraphics[angle=0, width=0.9\textwidth]{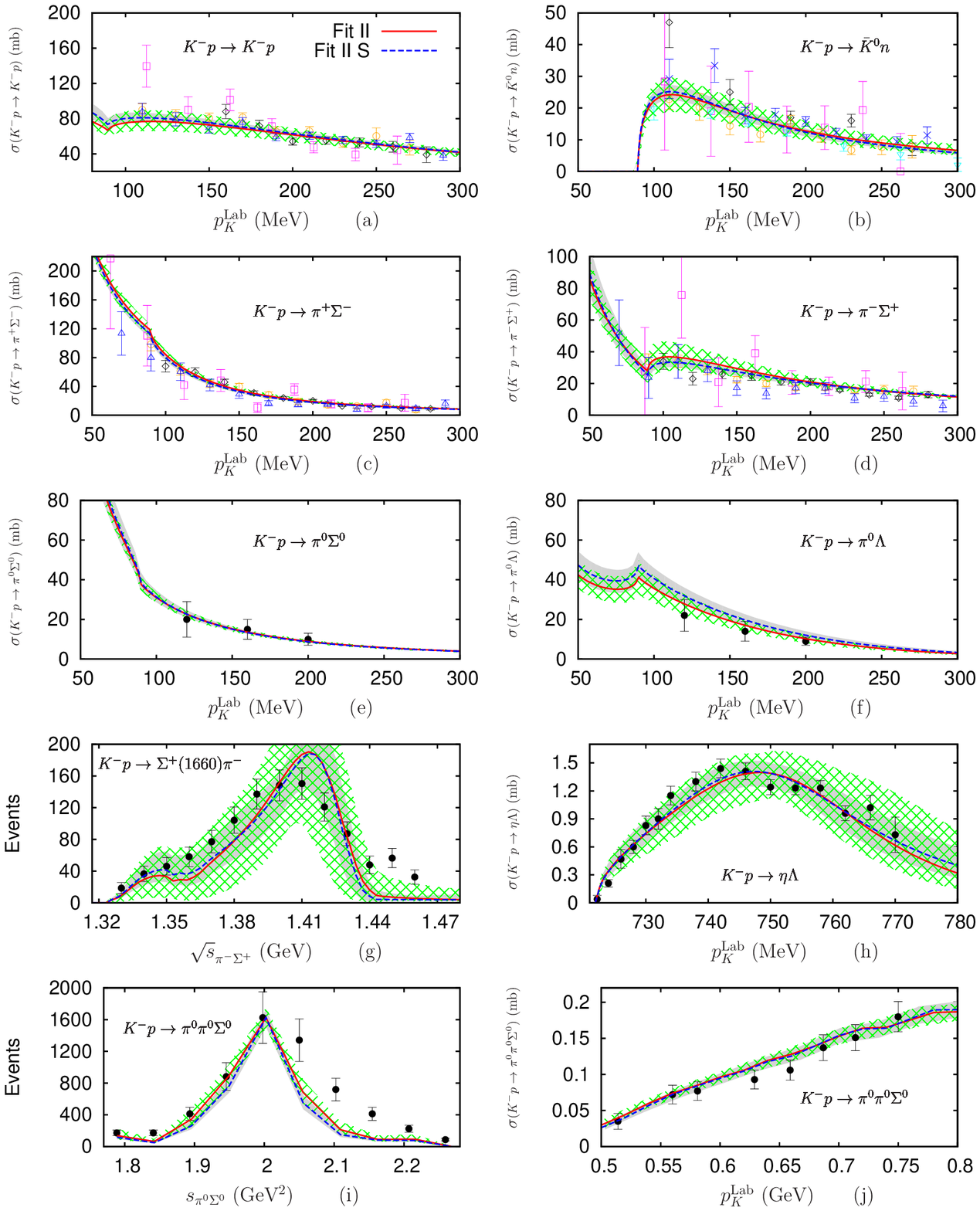}
\caption{{\small (Color online).  The same as in Fig.~\ref{fig1} but for Fit II. We refer to  Fig.~\ref{fig1} for notation.  }}
\label{figfit2cs}
\end{center}
\end{figure}

\begin{figure}[H]
\begin{center}
 \includegraphics[angle=0, width=0.9\textwidth]{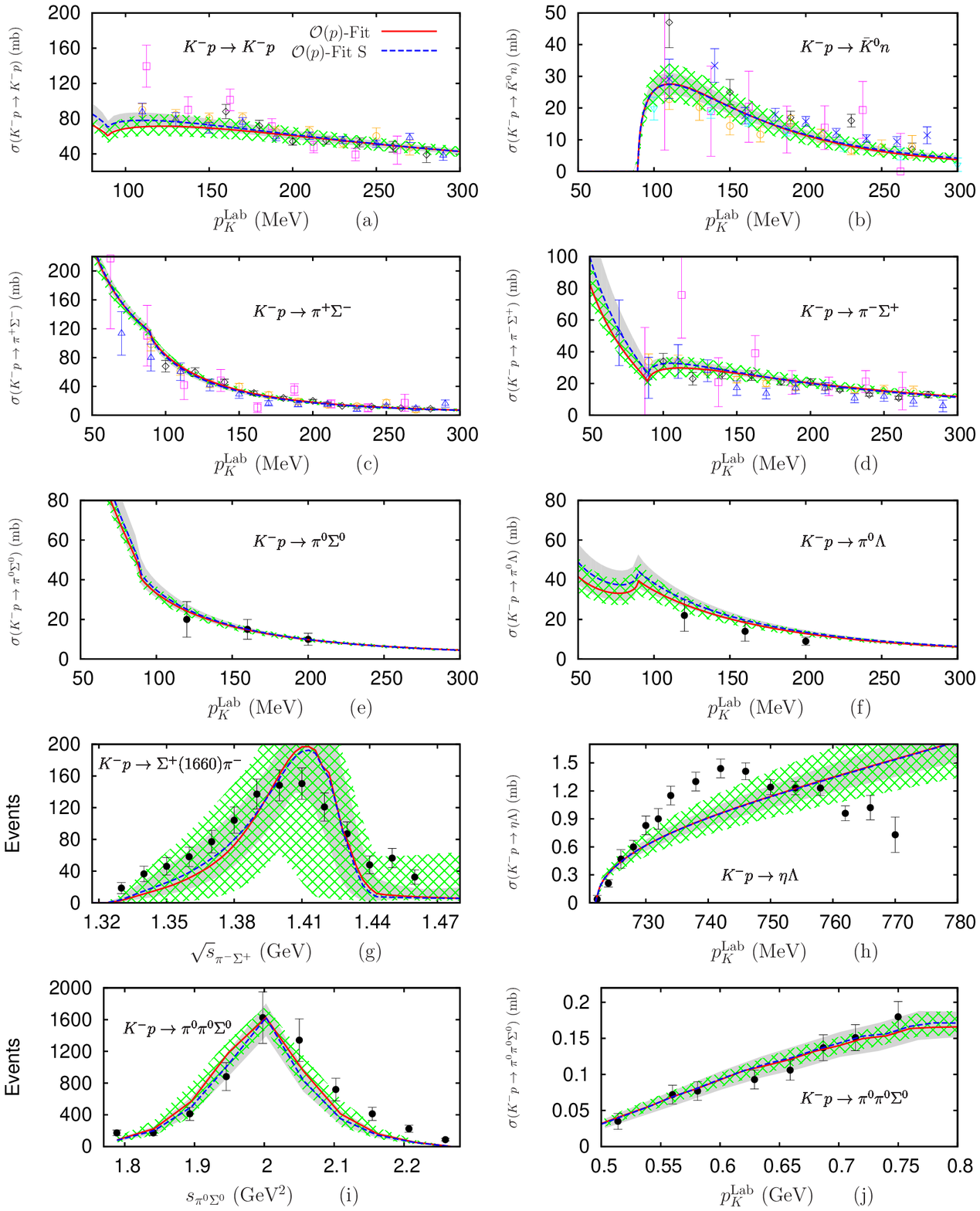}
\caption{{\small (Color online).  The same as in Fig.~\ref{fig1} but for the ${\cal O}(p)$-Fit. We refer to  Fig.~\ref{fig1} for notation.  }}
\label{figopcs}
\end{center}
\end{figure}

\begin{figure}[H]
\begin{minipage}[h]{0.49\textwidth}
\begin{center}
\includegraphics[angle=-0, width=1.0\textwidth]{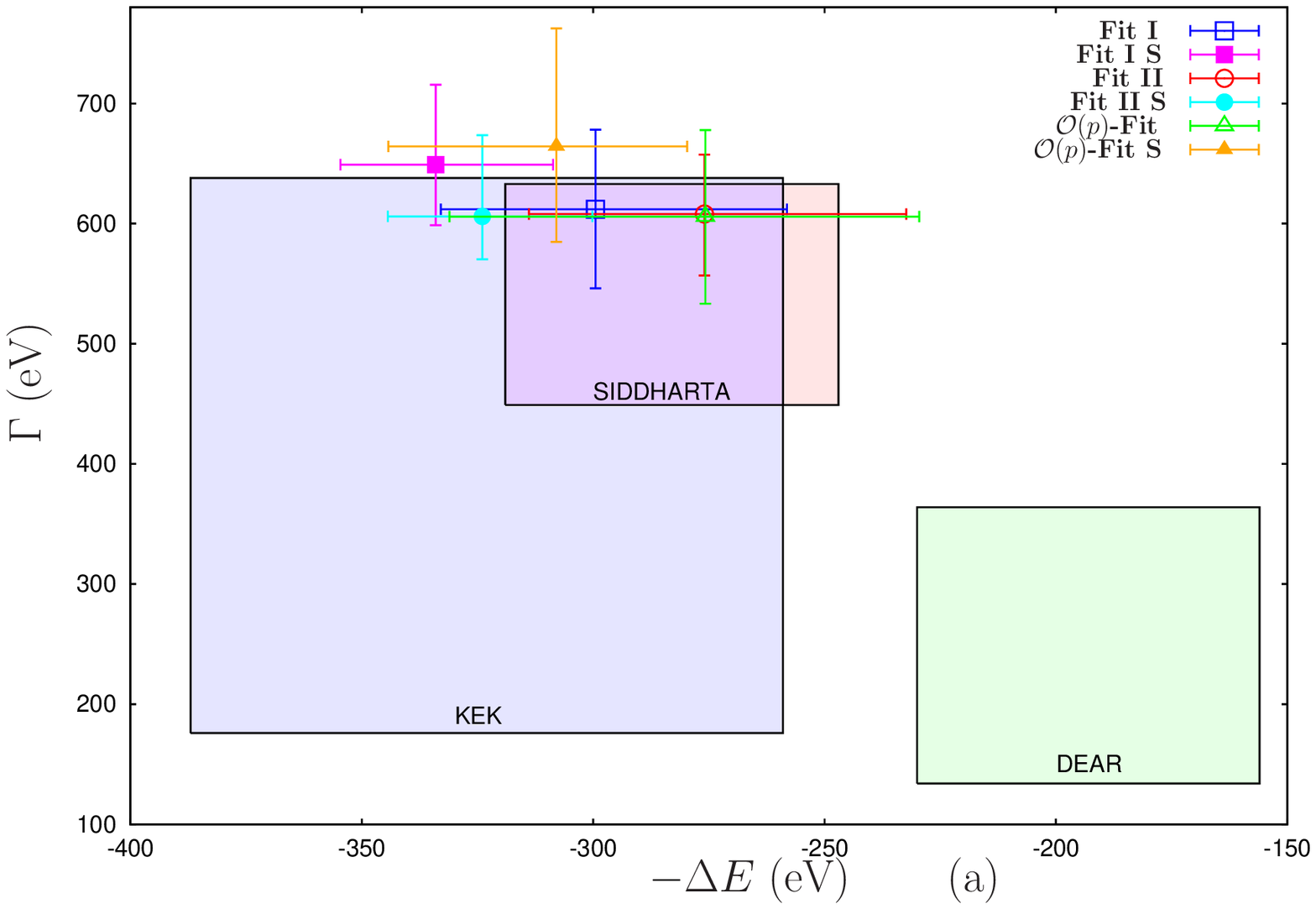}
\end{center}
\end{minipage}
\begin{minipage}[h]{0.49\textwidth}
\begin{center}
 \includegraphics[angle=-0, width=1.0\textwidth]{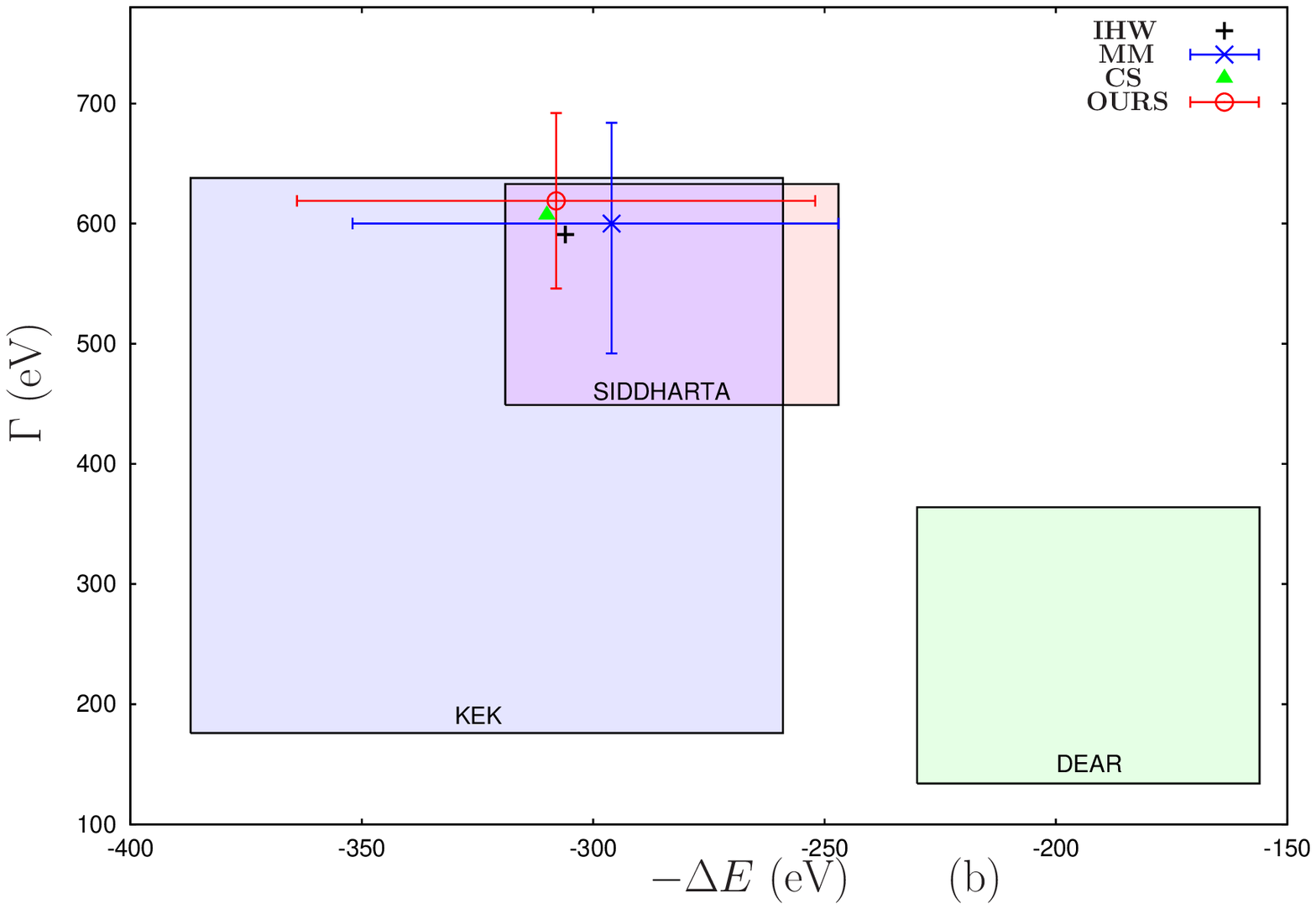}
\end{center}
\end{minipage}
\caption{{\small (Color online). Reversed sign energy shift $-\Delta E$ and width of the $1s$ state of kaonic hydrogen. The experimental 
data with errors indicated by the sides of the rectangles correspond to the KEK \cite{kek}, DEAR \cite{dear} and 
SIDDHARTA \cite{Bazzi:2011zj}. In the left side, i.e. in panel (a), the empty square, circle and triangle are our results from Table \ref{tabresult} for Fit I, Fit II and $\cO(p)$-Fit respectively,  while the filled square, circle and triangle correspond to the results from Table \ref{tabresultcomchisq} for Fit I, Fit II and $\cO(p)$-Fit, 
in order. In the right side, i.e. in panel (b), we show our final weight-averaged results in Table \ref{final_as_1s} by the (red) empty circle (denoted as OURS) and those from Ref.~\cite{Ikeda:2012au} (denoted as IHW in black plus symbol), Ref.~\cite{Cieply:2011nq} (denoted as CS in green triangle)  and Ref.~\cite{Mai:2012dt} (denoted as MM in blue cross).  } }
\label{fig:squares}
\end{figure}

\begin{figure}[H]
\begin{center}
 \includegraphics[angle=0, width=0.7\textwidth]{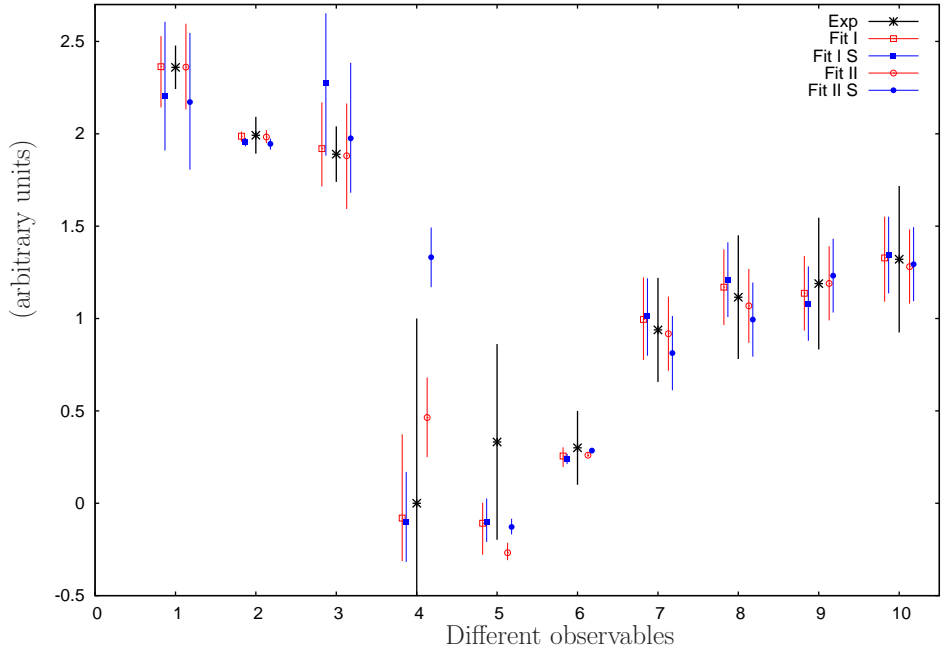}
\caption{{\small (Color online). We show from left to right  $\gamma$, $R_c$, $R_n$, $a_{0+}^+$, $\delta_{\pi \Lambda }$, 
$ \sigma_{\pi N}$ and the baryon masses for  $N $, $\Lambda$, $\Sigma$ and $\Xi$. 
For each observable the left most point corresponds to  Fit I, 
 next, going from left to right, one has  the experimental value and finally the result 
for Fit II. For every of the fits we show the results obtained by employing each definition of 
 $ \chi^2$, Eqs.~\eqref{defchisq} (empty symbols) and \eqref{defchisq2} (filled symbols), as indicated in the figure. 
Notice we have rescaled different observables in order to clearly show them in one plot. For the original values, one should see the details in the text. }
\label{figratios}}
\end{center}
\end{figure}

\begin{figure}[H]
\begin{center}
 \includegraphics[angle=0, width=0.8\textwidth]{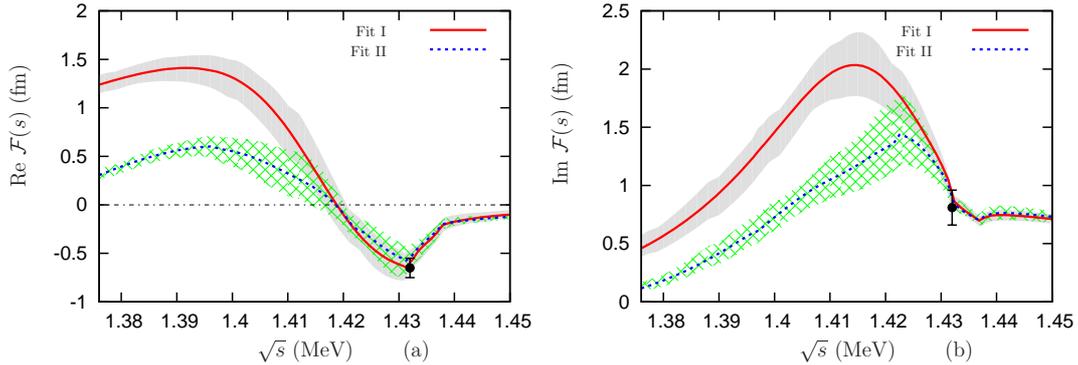}
\caption{{\small (Color online). Extrapolation of the $K^-p\to K^-p$ amplitude $\mathcal{F} (\sqrt{s})$ in Eq.~\eqref{defscattl} 
to the subthreshold energy region for Fit I (red solid line) and Fit II (dashed blue line) in Table~\ref{tabpara}. 
The real part is shown in the panel (a) and the imaginary part in the panel (b).
The shaded and hatched areas surrounding the solid and dashed lines, respectively,  correspond to the statistical error bands from the fit in Table~\ref{tabpara}. 
 The points with error bars correspond to the scattering length directly obtained from the SIDDHARTA measurement and the modified Deser-type formula, Eq.~\eqref{defegamma}.
} }
\label{fig2}
\end{center}
\end{figure}

\subsection{Results using the $\chi^2$ defined in Eq.~\eqref{defchisq2}}

Next, we consider the stability of our fits if the $\chi^2$ is defined according to the standard definition Eq.~\eqref{defchisq2} instead of using Eq.~\eqref{defchisq}. If  a big change is observed by doing this modification  in the outcome of a fit, then this clearly indicates that arbitrariness in the way chosen to describe nature is affecting our results. If this is the case 
 such a fit should be discounted. We conceive this check as a stability criterion. The fitted parameters for these new fits are shown in Table~\ref{tabparacomchisq}. At the semiquantitative level they are quite similar to those given in  Table~\ref{tabpara}. The only exception is $a_1$ for Fit I, although this parameter  appears unstable when changing from one fit to the other within the  same definition of $\chi^2$. This is already the case in Table~\ref{tabpara}. For the cross sections and event distributions shown in Figs.~\ref{fig1}, \ref{figfit2cs} and \ref{figopcs}, the new curves, given by the dashed lines, differ very little from the solid lines obtained previously. The uncertainty in the new curves is indicated by the shaded band around each dashed line. Then,  there is stability under the change in the definition of the $\chi^2$ for these observables.
 We now consider those magnitudes given in Table~\ref{tabresult}. The new values employing the standard definition Eq.~\eqref{defchisq2} 
are given in Table~\ref{tabresultcomchisq} and are shown pictorially in Figs.~\ref{fig:squares} and~\ref{figratios} by 
the filled symbols (squares, circles and triangles). Taking into account the 
error bars the results are compatible between the two definitions of the $\chi^2$. The only exception happens for $a_{0+}^+$ in Fit II, for which the value obtained employing the definition of the $\chi^2$ in Eq.~\eqref{defchisq2} is clearly different to the value obtained previously.  Nevertheless, since the systematic uncertainty for our calculation of $a_{0+}^+$ is big, as indicated by the 
large error bar attached to the cross in Fig.~\ref{figratios}, we consider that this deficiency is not significant.
 Regarding the ${\cal O}(p)$-Fit, not shown in Fig.~\ref{figratios}, one can observe by comparing the fifth columns in Tables~\ref{tabresult} and \ref{tabresultcomchisq} that the variation of all the quantities is within the one-sigma range.

 For the energy shift and width of the $1s$ state of kaonic hydrogen we see clearly in the left panel of Fig.\ref{fig:squares} that the new values, given by the filled square (Fit I), circle (Fit II) and triangle (${\cal O}(p)$-Fit) are within the estimated errors compatible with those obtained before employing the definition Eq.~\eqref{defchisq}. Nevertheless, the central values for each fit move around one sigma from each other when changing the definition of $\chi^2$. We also give in Table~\ref{table:scatcomchisq} the values of the scattering lengths $a_{K^-p}$, $a_{I=0}$ and $a_{I=1}$ for these new fits employing the definition Eq.~\eqref{defchisq2} for the $\chi^2$. They are perfectly compatible to those given before in Table~\ref{table:scat}. In summary, our stability criterion is well fulfilled. 
\begin{table}[ht]
\renewcommand{\tabcolsep}{0.07cm}
\renewcommand{\arraystretch}{1.3}
{\footnotesize 
\begin{center}
\begin{tabular}{|c|c|c|c|}
\hline
Observable & Fit I & Fit II & ${\cal O}(p)$-Fit \\
\hline
$a_{K^- p}$ (fm) & $-0.75_{-0.08}^{+0.07}+i\,1.00_{-0.08}^{+0.08}$  & $-0.74_{-0.08}^{+0.07}+i\,0.93_{-0.08}^{+0.09}$ & $-0.67_{-0.09}^{+0.08}+i\,1.00_{-0.12}^{+0.15}$ \\ 
\hline
$a_{I=0}$  (fm) & $-1.86_{-0.11}^{+0.14}+i\,1.35_{-0.15}^{+0.17}$  & $-1.79_{-0.14}^{+0.13}+i\,1.36_{-0.19}^{+0.18}$ & $-1.72_{-0.17}^{+0.18}+i\,1.47_{-0.23}^{+0.30}$ \\ 
\hline
$a_{I=1}$ (fm)  & $0.37_{-0.04}^{+0.05}+i\,0.65_{-0.06}^{+0.07}$   & $0.31_{-0.06}^{+0.05}+i\,0.50_{-0.05}^{+0.05}$ & $0.38_{-0.07}^{+0.05}+i\,0.52_{-0.11}^{+0.08}$ \\
\hline
\end{tabular}
\caption{
{\small $K^-p$ scattering length, $a_{K^-p}$, and $I=0$, $1$ $\bar{K}N$ scattering lengths, $a_{I=0}$, $a_{I=1}$, in order, for Fit I, Fit II and $\cO(p)$-Fit in Table~\ref{tabparacomchisq}.}
\label{table:scatcomchisq}
} 
\end{center}
}
\end{table}

From the discussion above, we take as resulting values from our study for the $\bar{K}N$ scattering lengths, $\Delta E$ and $\Gamma$  the mean and variance calculated from the four NLO fits. That is, Fit I and Fit II in Tables~\ref{tabpara} and \ref{tabparacomchisq}. For each quantity  we add in quadrature the variance obtained and the largest of the statistical errors resulting from the $\chi^2$ definitions Eq.~\eqref{defchisq2} and  Eq.~\eqref{defchisq}, this gives our final estimate for the error bar. In this way, the spread in the values by changing the $\chi^2$ definition is taken into account as a source of systematic uncertainty in our results given in Table~\ref{final_as_1s}. Our final values for $\Delta E$ and $\Gamma$ are also plotted in the right panel of Fig.~\ref{fig:squares} by the empty circle. There we also show some other recent determinations \cite{Ikeda:2011pi,Ikeda:2012au,Cieply:2011nq,Mai:2012dt} that include SIDDHARTA data \cite{Bazzi:2011zj} in their fits.

\begin{table}[ht]
\begin{center}
{\footnotesize
\begin{tabular}{|c|c|c|c|c|}
\hline
$\Delta E$ (eV) & $\Gamma$ (eV) & $a_{K^- p}$ (fm) & $a_{I=0}$ (fm) & $a_{I=1}$ (fm)\\ 
\hline
&&&&\\
 $308\pm 56$ & $619\pm73$ & $(-0.69\pm0.16) +i\,(0.94\pm0.11)$ 
&  $(-1.74\pm0.34) +i\,(1.31\pm0.20)$ &  $(0.36\pm0.12)+i\,(0.56\pm0.12)$ \\
\hline
\end{tabular}
}
\caption{{\small  Our final results for the quantities indicated in the first row. The error bars are calculated as 
explained in the text.  }
\label{final_as_1s}}
\end{center}
\end{table}

\section{Associated spectroscopy}
\label{poles}

Now we analyze the spectroscopy   by studying the pole content of the $S$-wave meson-baryon scattering amplitudes with strangeness $-1$ that result from the fits I and II. 
We discuss only the results obtained from the NLO 
fits in Table~\ref{tabpara}, employing the definition Eq.~\eqref{defchisq}, because those obtained from the NLO fits in Table~\ref{tabparacomchisq} are almost coincident.\footnote{We have explicitly checked that for the ${\cal O}(p)$-fit, except for the $\Lambda(1670)$ resonance that is not generated, the rest of isoscalar and isovector resonances are quite similar 
to those from the NLO fits and no further insight results.}  To perform this study, we need to 
extrapolate the meson-baryon scattering amplitudes from the physical or first Riemann sheet to the unphysical Riemann sheets 
in the complex energy plane. The physical Riemann sheet is such that the imaginary part of the modulus of the three-momentum associated with every channel is positive. The other Riemann sheets are defined depending on which three-momenta are evaluated in the other sheet of the square root, with an additional minus sign. In practice the change of sheet can be easily performed by adding a new 
term to $g(s)_i$, Eq.~\eqref{gs}, so that this function in its second Riemann sheet, $g(s)_{i;II}$, is given by \cite{Oller:1997ti}
\begin{align}
g(s)_{i;II}&=g(s)_{i}+ i\rho(s)~,\nn\\
\rho(s)&=\frac{q(s)_{i;I}}{4\pi W}~,
\end{align}
with $q(s)_{i;I}$ the CM three-momentum of the $i_{{\rm th}}$ state calculated in its first Riemann sheet, with positive imaginary part. Notice that in this way along the real axis and above the $i_{{\rm th}}$ threshold the imaginary part of $g(s)_{i;II}$ changes sign in the associated second Riemann sheet (denoted  by II) with respect to first Riemann sheet.
 Let us denote the physical sheet as $(+,+,+,+,\ldots)$, where we have ten entries inside the bracket corresponding to 
the sign of the imaginary part of every $q_i(s)$ function for the ten coupled channels. 
Strictly speaking there are 10 unphysical Riemann sheets that are connected continuously with the physical sheet by crossing 
from $s+ i 0^+$ to $s - i 0^+$, which can be symbolized as  $(-,+,+,+,\ldots)$, $(-,-,+,+,\ldots)$, $(-,-,-,+,\ldots)$ 
and so on, adding an additional minus sign every time a threshold is passed over. Nevertheless, we notice that the gaps between the three thresholds $\pi^0 \Sigma^0$, $\pi^- \Sigma^+$ 
and $\pi^+ \Sigma^-$ are quite narrow, and the same applies to the $K^- p$, $\bar{K}^0 n$ and $K^0 \Xi^0$, $K^+ \Xi^-$ channels. Indeed, every of the sets of thresholds indicated would be degenerate if isospin symmetry were not broken in the pseudoscalar and baryon masses.  In the following discussions, we do not distinguish the small differences of thresholds inside the three clusters. In such a way, we only have six unphysical Riemann sheets that are directly 
connected to the first one. We denote the second Riemann sheet as $(-,+,+,+,+,+,+,+,+,+)$, 
meaning that the sign of the imaginary part of $q_1$ is changed and becomes negative. 
The $3_{{\rm rd}}$ (3RS), $4_{{\rm th}}$ (4RS), $5_{{\rm th}}$ (5RS), $6_{{\rm th}}$ (6RS) and  $7_{{\rm th}}$ (7RS) Riemann sheets correspond 
to $(-,-,-,-,+,+,+,+,+,+)$, $(-,-,-,-,-,-,+,+,+,+)$, $(-,-,-,-,-,-,-,+,+,+)$, $(-,-,-,-,-,-,-,-,+,+)$ and $(-,-,-,-,-,-,-,-,-,-)$, respectively. Apart from the pole position $s_R$ in the complex energy plane, the resonance $R$ is also characterized by its 
couplings, which correspond to the residues $\beta_{i}$ of the resonance pole and are calculated through
\begin{equation}
 T_{ij} = - \lim_{s\to s_R}\frac{\beta_i \beta_j}{s-s_R}\,. 
\end{equation}
We summarize the resonance pole positions and their couplings in 
Table~\ref{tabpole1} for Fit I and Table~\ref{tabpole2} for Fit II.

 The two relevant resonances with $I=0$,  
namely $\Lambda(1405)$ and $\Lambda(1670)$, clearly show up in both fits
  The two poles corresponding to the $\Lambda(1405)$ are found in the 3RS for both Fit I and Fit II. 
The narrower poles, i.e. those with small imaginary parts, from the two fits are perfectly consistent,  
while  the broader ones differ slightly (taking into account errors) from the two fits. 
 The narrower poles from our study also agree well with the recent determinations  
from Refs.~\cite{Ikeda:2011pi,Ikeda:2012au,Oller:2006jw}. While for the broader poles, our results from 
Fit II are compatible with those in Refs.~\cite{Ikeda:2011pi,Ikeda:2012au}, that also use  the physical decay constants 
for $\pi$, $K$ and $\eta$ as in our Fit II. Concerning the broader pole, none of our results seem compatible with that in Ref.~\cite{Mai:2012dt}, 
though we find that the real part of the broader pole in the case of Fit I is slightly above the $K^-p$ threshold. 
As pointed out in Ref.~\cite{Borasoy:2005ie} the broader  pole is much more dependent on the details of the theoretical approach than 
the narrower one. Our results for the broader pole are well inside the interval of values obtained in the same reference. Despite the fact that the broader pole changes its pole position according to the fit considered, the $\pi\Sigma$ and $\bar{K}N$ amplitudes have their peak values at nearly the same energy. This is due to the fact that the gradient of the modulus of the amplitudes is tilted so that it is not perpendicular to the real axis from the pole positions. This is clearly shown by the contour plots in  Fig.~\ref{fig:contouri0res} where the modulus of the elastic $I=0$ $\bar{K}N$ (blue dashed) and $\pi\Sigma$ (red solid) $S$-waves for Fit I (left panel) and Fit II (right panel) are plotted. For Fit I the gradient is tilted to the left from the broader pole position while for Fit II it is tilted to the right from the analogous pole, being always oriented towards the narrower pole. However, the gradient is always tilted to the left of the narrower pole. This figure also shows that for  the $\pi\Sigma$ 
state the two 
poles play a role in its shape on the real axis while  $\bar{K}N$ is dominated by the narrower pole. In addition, the change in the position of the broader pole has more influence below the $\bar{K}N$ threshold and this is the origin of the differences between Fit I and Fit II in Fig.~\ref{fig2}, as can be easily inferred from Fig.~\ref{fig:contouri0res} considering the dashed lines.

\begin{figure}[H]
\begin{minipage}[h]{0.49\textwidth}
\begin{center}
\includegraphics[angle=-0, width=1.0\textwidth]{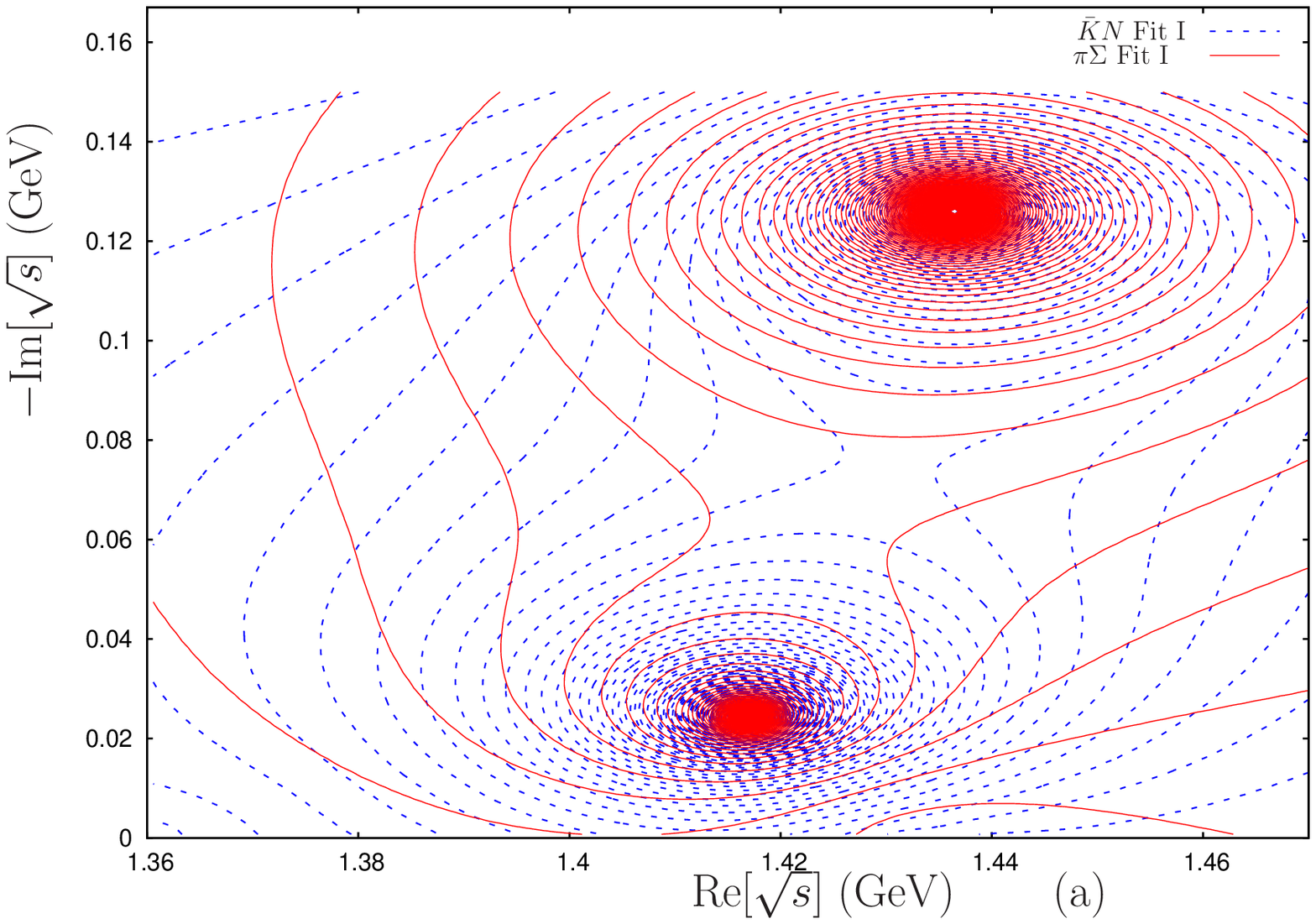}
\end{center}
\end{minipage}
\begin{minipage}[h]{0.49\textwidth}
\begin{center}
 \includegraphics[angle=-0, width=1.0\textwidth]{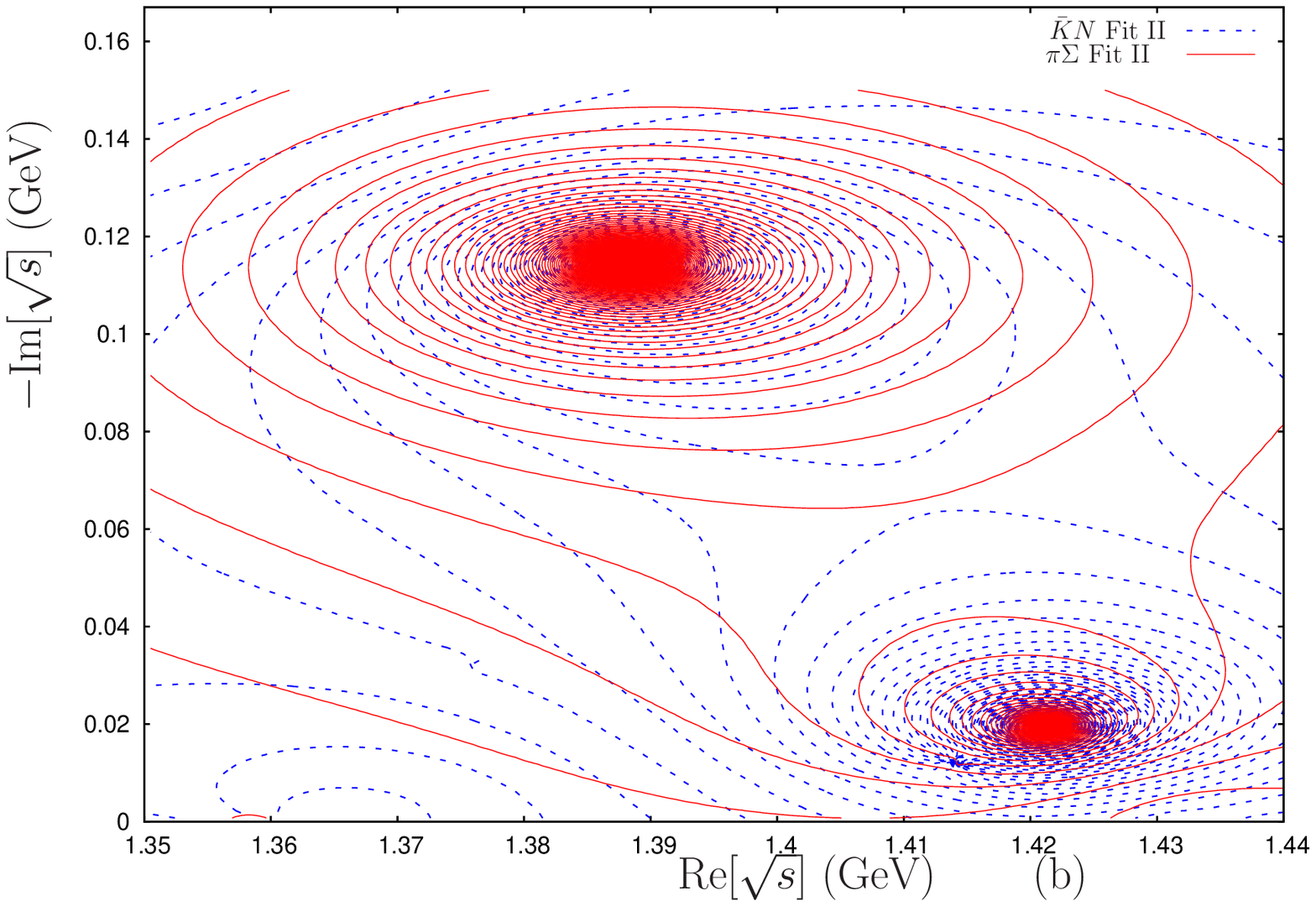}
\end{center}
\end{minipage}
\caption{{\small (Color online). Contour plot for the modulus of the elastic $I=0$ $\bar{K}N$ (blue dashed) and $\pi\Sigma$ (red solid) $S$-waves for Fit I [panel (a)]  
and Fit II [panel (b)].  } }
\label{fig:contouri0res}
\end{figure}

In addition to the pole positions, we also calculate the couplings of the resonances, 
see Tables~\ref{tabpole1} and \ref{tabpole2}. 
The first conclusion we can make for the $\Lambda(1405)$ is that both the narrower and broader 
poles couple  mostly to the $\pi\Sigma$ and $\bar{K}N$ channels. 
For the broader poles, the coupling strength to $\pi\Sigma$ is around twice as large as  the one for the narrower pole, 
while the coupling to $\bar{K}N$ for both the narrower and broader poles is quite the same. 
Taking into account that the narrower pole gives strength around the $\bar{K}N$ threshold it  dominates a reaction with production of $\bar{K}N$, 
as pointed out above regarding Fig.~\ref{fig:contouri0res}. In turn, the signal for $\pi\Sigma$ is more dominated by the broader pole 
that  couples  more strongly to this channel than the other narrower and higher in mass pole, see also     Fig.~\ref{fig:contouri0res}. 
 Concerning the channels with thresholds larger than the resonance masses, such as $\eta\Lambda$ and $K\Xi$, 
 the narrower and broader poles couple to them weakly, much less than to $\pi\Sigma$ and $\bar{K}N$. 

Next we consider the $\Lambda(1670)$ resonance. The properties of this resonance are mainly determined by 
the $K^-p\to\eta\Lambda$ reaction that is included in our fits \cite{starostin01prc}, but not in the recent studies 
of Refs.~\cite{Ikeda:2011pi,Ikeda:2012au,Cieply:2011nq,Mai:2012dt}. As a result these references do not obtain any 
 information about the $\Lambda(1670)$ resonance. In our case, Fit I and Fit II lead to quite similar, good, reproductions 
 for the $K^-p\to\eta\Lambda$ cross section, see the eighth panels in Figs.~\ref{fig1} and \ref{figfit2cs}, respectively. 
As a result, the $\Lambda(1670)$ pole positions and couplings from both fits perfectly agree between each other. They also agree with those of Ref.~\cite{Oller:2006jw,Liu:2012} and with the PDG results \cite{pdg}. 
Very similar poles in three different Riemann sheets, namely 3RS, 4RS and 5RS, are found. As explained in Ref.~\cite{Oller:2006jw,Au:1986vs} 
all these poles reflect the same underlying resonance since they are connected continuously when changing 
the different Riemann sheets indicated in a soft way via a continuous parameter.
 We conclude that the $\Lambda(1670)$ is most strongly coupled to the $K\Xi$ channel and has similar coupling strengths 
to $\pi\Sigma, \bar{K}N$ and $\eta\Lambda$ channels. These couplings are also quite similar to those found in Ref.~\cite{Oller:2006jw}.

Regarding the $I=1$ poles the resonance content is less clear since it depends on the details of the fits performed. 
 For both NLO fits one has broad poles with central value masses of 1646 and 1686~MeV for fits I and II, respectively, that are in correspondence 
with the properties of the bumps associated with the $\Sigma(1620)$ in the PDG \cite{pdg}. 
 Additionally, Fit II shows a pole at the central value position of $1741-i\,94$~MeV with mass and width in the range 
given by the PDG \cite{pdg} for the $\Sigma(1750)$ resonance. From the residues shown in Table~\ref{tabpole2} its main decay channels are 
$\pi\Sigma$, $\bar{K}N$ and $\eta \Sigma$ as in \cite{pdg}. Nonetheless,  in our case the $\eta\Sigma$ decay width is smaller than the value suggested 
in \cite{pdg} with large errors. The two last poles in Table~\ref{tabpole2} with $I=1$  are connected in a continuous way by changing from the 5RS to the 6RS with a continuous parameter so that  both poles correspond to the same 
physical object seen in different Riemann sheets \cite{Oller:2006jw,Au:1986vs}.  In Fit II we also have  narrow poles near the $\bar{K}N$ threshold. 
Similar poles were also reported in Refs.~\cite{Oller:2000fj,Oller:2006jw}.

\begin{table}[ht]
\renewcommand{\tabcolsep}{0.08cm}
\renewcommand{\arraystretch}{1.4}
{\footnotesize
\begin{center}
\begin{tabular}{|c|c|c|c|c|c|c|c|c|c|c|c|c|}
\hline
Pole & $|\beta_{\pi\Lambda}|$ & $|\beta_{\pi\Sigma}|_0$ & $|\beta_{\pi\Sigma}|_1$ 
 & $|\beta_{\pi\Sigma}|_2$ & $|\beta_{\bar{K} N}|_0$ & $|\beta_{\bar{K} N}|_1$ & $|\beta_{\eta\Lambda}|$ 
 & $|\beta_{\eta\Sigma}|$ & $|\beta_{K\Xi}|_0$ & $|\beta_{K\Xi}|_1$\\
\hline
$\Lambda(1405)$ &  &   &   &  &  &   &  &   &   &  \\ 
$1436_{-10}^{+14}-i\,126_{-28}^{+24}$ (3RS) & 0.0$_{-0.0}^{+0.0}$ & 8.8$_{-0.4}^{+0.9}$ & 0.0$_{-0.0}^{+0.0}$ 
& 0.0$_{-0.0}^{+0.0}$ & 7.7$_{-0.7}^{+1.3}$ & 0.0$_{-0.0}^{+0.1}$ & 1.4$_{-0.3}^{+0.4}$ & 0.0$_{-0.0}^{+0.1}$ 
& 2.1$_{-0.7}^{+0.8}$ & 0.0$_{-0.0}^{+0.0}$ \\
$1417_{-4}^{+4}-i\,24_{-4}^{+7}$  (3RS) & 0.1$_{-0.0}^{+0.0}$ & 5.0$_{-0.8}^{+1.5}$ & 0.1$_{-0.0}^{+0.0}$ 
& 0.0$_{-0.0}^{+0.0}$ & 7.7$_{-0.6}^{+1.2}$ & 0.1$_{-0.0}^{+0.0}$ & 1.4$_{-0.3}^{+0.4}$ & 0.1$_{-0.0}^{+0.0}$ 
& 1.5$_{-0.5}^{+0.7}$ & 0.1$_{-0.0}^{+0.0}$ \\ 
\hline
$\Lambda(1670)$ &  &   &   &  &  &   &  &   &   &  \\ 
$1674_{-2}^{+3}-i\,8_{-3}^{+6}$  (4RS) & 0.0$_{-0.0}^{+0.0}$ & 0.8$_{-0.1}^{+0.4}$ & 0.0$_{-0.0}^{+0.0}$ 
& 0.0$_{-0.0}^{+0.0}$ & 1.5$_{-0.2}^{+0.4}$ & 0.0$_{-0.0}^{+0.0}$ & 1.5$_{-0.2}^{+0.2}$ & 0.0$_{-0.0}^{+0.0}$ 
& 10.8$_{-0.2}^{+0.2}$ & 0.1$_{-0.0}^{+0.0}$ \\
$1674_{-3}^{+3}-i\,11_{-3}^{+7}$ (5RS) & 0.0$_{-0.0}^{+0.0}$ & 0.9$_{-0.2}^{+0.4}$ & 0.0$_{-0.0}^{+0.0}$ 
& 0.0$_{-0.0}^{+0.0}$ & 1.6$_{-0.2}^{+0.4}$ & 0.0$_{-0.0}^{+0.0}$ & 1.7$_{-0.3}^{+0.5}$ & 0.0$_{-0.0}^{+0.0}$ 
& 11.1$_{-0.3}^{+0.3}$ & 0.1$_{-0.0}^{+0.0}$ \\
$1673_{-3}^{+3}-i\,11_{-3}^{+7}$ (6RS) & 0.0$_{-0.0}^{+0.0}$ & 0.9$_{-0.2}^{+0.4}$ & 0.0$_{-0.0}^{+0.0}$ 
& 0.0$_{-0.0}^{+0.0}$ & 1.6$_{-0.2}^{+0.4}$ & 0.0$_{-0.0}^{+0.0}$ & 1.7$_{-0.3}^{+0.5}$ & 0.0$_{-0.0}^{+0.0}$ 
& 11.1$_{-0.3}^{+0.3}$ & 0.1$_{-0.0}^{+0.0}$ \\
\hline
$\Sigma$~$I=1$ &  &   &   &  &  &   &  &   &   &  \\ 
$1646_{-127}^{+30}-i\,160_{-36}^{+78}$ (4RS,5RS) & 3.1$_{-0.5}^{+1.4}$ & 0.0$_{-0.0}^{+0.0}$ & 3.0$_{-0.5}^{+0.4}$ 
& 0.0$_{-0.0}^{+0.0}$ & 0.0$_{-0.0}^{+0.0}$ & 2.9$_{-0.3}^{+0.4}$ & 0.0$_{-0.0}^{+0.0}$ & 7.9$_{-1.2}^{+1.1}$ 
& 0.0$_{-0.0}^{+0.0}$ & 6.4$_{-2.2}^{+1.4}$ \\ 
$1878_{-59}^{+48}-i\,169_{-34}^{+27}$ (6RS) & 1.0$_{-0.4}^{+0.5}$ & 0.0$_{-0.0}^{+0.0}$ & 5.8$_{-0.6}^{+0.9}$ 
& 0.0$_{-0.0}^{+0.0}$ & 0.0$_{-0.0}^{+0.0}$ & 3.7$_{-0.4}^{+0.4}$ & 0.0$_{-0.0}^{+0.0}$ & 3.9$_{-1.2}^{+1.1}$ 
& 0.1$_{-0.0}^{+0.0}$ & 16.1$_{-1.6}^{+2.4}$ \\ 
\hline
\end{tabular}
\caption{{\small Resonances from Fit I in Table \ref{tabpara}. The positions of the resonance poles are in MeV and 
modulus of their couplings are given in GeV.  The latter are represented by $|\beta_i|_{(I)}$, where the isospin label $I$ is indicated when more than one isospin channel is possible.} 
\label{tabpole1}}
\end{center}
}
{\footnotesize
\begin{center}
\begin{tabular}{|c|c|c|c|c|c|c|c|c|c|c|c|c|}
\hline
Pole & $|\beta_{\pi\Lambda}|$ & $|\beta_{\pi\Sigma}|_0$ & $|\beta_{\pi\Sigma}|_1$ 
 & $|\beta_{\pi\Sigma}|_2$ & $|\beta_{\bar{K} N}|_0$ & $|\beta_{\bar{K} N}|_1$ & $|\beta_{\eta\Lambda}|$ 
 & $|\beta_{\eta\Sigma}|$ & $|\beta_{K\Xi}|_0$ & $|\beta_{K\Xi}|_1$\\
\hline
$\Lambda(1405)$ &  &   &   &  &  &   &  &   &   &  \\ 
$1388_{-9}^{+9}-i\,114_{-25}^{+24}$ (3RS) & 0.0$_{-0.0}^{+0.0}$ & 8.2$_{-0.5}^{+0.8}$ & 0.0$_{-0.0}^{+0.0}$ 
& 0.0$_{-0.0}^{+0.0}$ & 6.1$_{-0.6}^{+1.1}$ 
& 0.1$_{-0.0}^{+0.0}$ & 2.2$_{-0.3}^{+0.6}$ & 0.0$_{-0.0}^{+0.0}$ & 1.9$_{-0.1}^{+0.2}$ & 0.1$_{-0.0}^{+0.0}$ \\
$1421_{-2}^{+3}-i\,19_{-5}^{+8}$  (3RS) & 0.2$_{-0.1}^{+0.1}$ & 4.2$_{-0.9}^{+1.5}$ 
& 0.2$_{-0.0}^{+0.0}$ & 0.0$_{-0.0}^{+0.0}$ 
& 6.2$_{-0.5}^{+1.2}$ & 0.3$_{-0.1}^{+0.1}$ & 2.8$_{-0.3}^{+0.5}$ & 0.4$_{-0.1}^{+0.2}$ & 0.7$_{-0.3}^{+0.4}$ 
& 0.4$_{-0.1}^{+0.1}$ \\ 
\hline
$\Lambda(1670)$ &  &   &   &  &  &   &  &   &   &  \\ 
$1676_{-3}^{+5}-i\,7_{-3}^{+5}$  (4RS) & 0.0$_{-0.0}^{+0.0}$ & 0.9$_{-0.1}^{+0.1}$ & 0.0$_{-0.0}^{+0.0}$ 
& 0.0$_{-0.0}^{+0.0}$ & 1.5$_{-0.4}^{+0.4}$ & 0.1$_{-0.0}^{+0.0}$ & 1.6$_{-0.2}^{+0.2}$ & 0.1$_{-0.0}^{+0.0}$ 
& 10.0$_{-0.1}^{+0.1}$ & 0.1$_{-0.0}^{+0.0}$ \\
$1677_{-3}^{+5}-i\,11_{-3}^{+5}$ (5RS) & 0.0$_{-0.0}^{+0.0}$ & 0.8$_{-0.1}^{+0.1}$ & 0.1$_{-0.0}^{+0.0}$ 
& 0.0$_{-0.0}^{+0.0}$ & 1.6$_{-0.4}^{+0.4}$ & 0.1$_{-0.0}^{+0.0}$ & 1.8$_{-0.2}^{+0.2}$ & 0.1$_{-0.0}^{+0.0}$ 
& 10.5$_{-0.2}^{+0.2}$ & 0.1$_{-0.0}^{+0.0}$ \\
$1677_{-3}^{+5}-i\,11_{-3}^{+5}$ (6RS) & 0.0$_{-0.0}^{+0.0}$ & 0.8$_{-0.1}^{+0.1}$ & 0.0$_{-0.0}^{+0.0}$ 
& 0.0$_{-0.0}^{+0.0}$ & 1.6$_{-0.4}^{+0.4}$ & 0.0$_{-0.0}^{+0.0}$ & 1.8$_{-0.2}^{+0.2}$ & 0.1$_{-0.0}^{+0.0}$ 
& 10.5$_{-0.2}^{+0.2}$ & 0.0$_{-0.0}^{+0.0}$ \\
\hline
$\Sigma$~$I=1$ &  &   &   &  &  &   &  &   &   &  \\ 
$1376_{-3}^{+3}-i\,33_{-5}^{+5}$  (3RS) & 2.0$_{-0.1}^{+0.1}$ & 0.0$_{-0.0}^{+0.0}$ & 0.1$_{-0.1}^{+0.1}$ 
& 0.0$_{-0.0}^{+0.0}$ & 0.1$_{-0.0}^{+0.0}$ & 2.1$_{-0.4}^{+0.5}$ & 0.0$_{-0.0}^{+0.0}$ 
& 4.0$_{-0.3}^{+0.5}$ & 0.0$_{-0.0}^{+0.0}$ & 6.3$_{-0.2}^{+0.2}$ \\ 
$1414_{-3}^{+2}-i\,12_{-2}^{+1}$  (3RS) & 1.9$_{-0.1}^{+0.1}$ & 0.3$_{-0.1}^{+0.1}$ & 1.0$_{-0.1}^{+0.2}$ 
& 0.0$_{-0.0}^{+0.0}$ & 0.4$_{-0.1}^{+0.2}$ & 2.5$_{-0.4}^{+0.3}$ & 0.2$_{-0.1}^{+0.1}$ 
& 3.3$_{-0.4}^{+0.4}$ & 0.1$_{-0.0}^{+0.0}$ & 3.3$_{-0.3}^{+0.3}$ \\
$1686_{-18}^{+18}-i\,101_{-8}^{+9}$ (5RS) & 0.2$_{-0.1}^{+0.1}$ & 0.0$_{-0.0}^{+0.0}$ & 3.5$_{-0.2}^{+0.2}$ 
& 0.0$_{-0.0}^{+0.0}$ & 0.0$_{-0.0}^{+0.0}$ & 3.5$_{-0.1}^{+0.1}$ & 0.0$_{-0.0}^{+0.0}$ & 3.9$_{-0.3}^{+0.3}$ 
& 0.1$_{-0.0}^{+0.0}$ & 10.9$_{-0.2}^{+0.2}$ \\ 
$1741_{-13}^{+12}-i\,94_{-3}^{+3}$  (6RS) & 1.1$_{-0.1}^{+0.1}$ & 0.0$_{-0.0}^{+0.0}$ & 2.3$_{-0.1}^{+0.1}$ 
& 0.0$_{-0.0}^{+0.0}$ & 0.0$_{-0.0}^{+0.0}$ & 2.8$_{-0.1}^{+0.1}$ & 0.0$_{-0.0}^{+0.0}$ & 3.7$_{-0.2}^{+0.2}$ 
& 0.1$_{-0.0}^{+0.0}$ & 7.9$_{-0.2}^{+0.3}$ \\ 
\hline
\end{tabular}
\caption{{\small Resonances from Fit II in Table \ref{tabpara}. For notation see Table~\ref{tabpole1}.}
\label{tabpole2}}
\end{center}
}
\end{table}

\section{Conclusions}
\label{conclusions}

 We have studied the strangeness $-1$ $S$-wave meson-baryon scattering with 10 coupled channels by applying Unitary  ChPT (a unitarization method based on an approximate algebraic solution to the N/D method) with the interaction kernel calculated 
up to $\mathcal{O}(p^2)$. The latter then corresponds to the NLO ChPT meson-baryon partial wave amplitudes. This study is prompted by the new precise measurement of the energy shift and width of the ground state of kaonic hydrogen. 
 We have successfully reproduced 
a large amount of experimental data, which include the cross sections of 
$K^-p\to \big\{ K^-p$, $\bar{K}^0n$, $\pi^+\Sigma^-$, $\pi^-\Sigma^+$, $\pi^0\Sigma^0$, $\pi^0\Lambda$, $\eta\Lambda$,
$\pi^0\pi^0\Sigma^0\big\}$,  $\pi^-\Sigma^+$ and $\pi^0\Sigma^0$ event distributions, three branching ratios in Eq.~\eqref{ratios} 
measured at the $K^-p$ threshold, the $\pi\Lambda$ phase shift measured at the $\Xi^-$ mass and the
 new SIDDHARTA measurement on the energy shift and width of the kaonic hydrogen $1s$ state Eq.~\eqref{defegamma}. We constrain 
further the fits by including, with rather large error bars, the pion-nucleon scalar isoscalar $S$-wave scattering length, the nucleon $\sigma$ term and the masses of $N, \Lambda, \Sigma$ and $\Xi$. We confirm the consistency between scattering data and the SIDDHARTA measurement, already 
highlighted in Refs.~\cite{Ikeda:2011pi,Ikeda:2012au,Cieply:2011nq,Mai:2012dt}, although we consider additional recent data from the reactions $K^-p\to \eta\Lambda$ \cite{starostin01prc} and 
 $K^-p\to \pi^0\pi^0\Sigma^0$ \cite{prakhov04prc}, not included in those references.
 We have performed two types of fits. In Fit I 
we employ a common pseudoscalar  decay constant, while in Fit II we distinguish the $\pi$, $K$ and $\eta$ weak decay constants. These two strategies 
are consistent with our ${\cal O}(p^2)$ calculation of the interaction kernel since the difference between weak decay constants 
gives rise to higher orders in the counting. In addition we have studied for every fit the two definitions for the $\chi^2$ typically used in the literature for 
the study of strangeness $-1$ $S$-wave scattering. In one of them more weight is given to the presumably precise measurements of the energy shift and width of the $1s$ state of  
kaonic hydrogen. We observe that the results are quite stable under the change of the definition for the $\chi^2$.
On the other hand,  despite the knowledge of the $K^-p$ scattering length is an important ingredient to constrain the subthreshold extrapolation of the 
$K^-p$ $S$-wave scattering amplitude, it is not enough to determine it in a precise manner. 
This  is clear by the significant differences in the subthreshold extrapolations resulting from Fit I and Fit II. 
In this sense, the systematic uncertainty affecting such extrapolation is larger than the statistical one reported 
in Refs.~\cite{Ikeda:2011pi,Ikeda:2012au,Cieply:2011nq,Mai:2012dt}. 
The situation might be improved in the future once the interaction kernel is calculated at ${\cal O}(p^3)$ in Unitary ChPT. 

 Discussions on the baryon resonance spectroscopy are carried out as well, 
 for which we calculate both the resonance pole positions and their residues (or resonance couplings). 
We find that the properties for the relevant isoscalar 
resonances are quite robust from the two fits. The two-pole structure for the $\Lambda(1405)$ is confirmed, in which the properties of the narrower pole, close to the $\bar{K}N$ threshold,  are quite stable. 
Our results are consistent with those of Refs.~\cite{Ikeda:2011pi,Ikeda:2012au,Oller:2006jw,Borasoy:2005ie}. 
For the other broader poles, somewhat small differences 
from the two fits are observed,  with one pole above and the other below the $K^-p$ threshold. These broader poles depend more on the details of the theoretical approach and on the fit performed. Despite that, the peak positions  on the real energy axis  for the $\bar{K}N$ and $\pi\Sigma$ amplitudes are quite the same in both fits, because the different orientation of the gradient of the modulus of the amplitudes with respect to the broader pole position. The gradients always point towards the position of the narrower pole much closer to the energy axis. Nevertheless, a precise knowledge of the pole position of the broader pole is required in order to end with a precise extrapolation of the elastic $K^-p$ scattering amplitude below the $\bar{K}N$ threshold.  The positions of the pole and couplings for the $\Lambda(1670)$ from the two fits are nearly identical and in good agreement with the 
properties reported in the PDG. The isovector resonances depend more on the details of the fit. Nevertheless, in both cases one observes broad poles in the region of the  $\Sigma(1620)$ resonance. Fit II also reports a pole position that could be identified with the resonance $\Sigma(1750)$. This fit accumulates strength around the $\bar{K}N$ threshold in $I=1$ too, corresponding to two relatively narrow poles in this energy region.

\section*{Acknowledgements}
This work is partially funded by the grants MEC  FPA2010-17806, the Fundaci\'on S\'eneca 11871/PI/09, 
the BMBF grant 06BN411, the EU-Research Infrastructure
Integrating Activity ``Study of Strongly Interacting Matter" (HadronPhysics2, grant No. 227431)
under the Seventh Framework Program of EU and the Consolider-Ingenio 2010 Programme CPAN (CSD2007-00042).
Z.H.G. also acknowledges the grants National Natural Science Foundation of China (NSFC) under contract No. 11105038, 
Natural Science Foundation of Hebei Province with contract No. A2011205093 and Doctor Foundation of Hebei Normal 
University with contract No. L2010B04.

\end{document}